# Flow-polarity decoupling and overall mobility enhancement in dense bacterial active fluids

Yuhao Wang [1], Premkumar Leishangthem [2], Yiming Ding [1], Xinliang Xu [2], Yilin Wu [1]

[1] *Department of Physics and Shenzhen Research Institute, The Chinese University of Hong Kong, Shatin, NT, Hong Kong, P.R. China.*
[2] School of Physics and Optoelectronic Engineering, Hainan University, Haikou, 570228, P. R. China

## Abstract
Active fluids consisting of living cells or synthetic microswimmers display rich emergent behavior and nonequilibrium mechanical properties, which not only shed light on various biological processes but also inform the engineering of autonomous fluidics and self-driven materials. The individual behavior of microswimmers and their interaction with self-generated mesoscale solvent flows underlie the emergent properties of active fluids. Here we studied the microscopic dynamics in dense 3D bacterial active fluids by simultaneous imaging of cell body, flagella, and flow field. A surprising finding is that the polarity of cells was randomly distributed in mesoscale flow regimes, and yet the system displays mesoscale order in the self-generated solvent flows. Despite the apparent flow-polarity decoupling, the motion of cells relative to local solvent flows predominantly navigated upstream, with the self-advection speed overall enhanced by a flow-controlled constant. Numerical modeling with full hydrodynamic interactions reveals that the observed flow-polarity decoupling arises from the breakdown of the commonly held force-dipole assumption for anisotropic microswimmers: in the presence of flow gradient and near-field hydrodynamic interactions, the direction of total active forcing exerted by a swimming bacterium to the surrounding fluid no longer aligns with its polarity. The simulations suggest that near-field interactions serve as a new type of emergent, configuration-dependent active forcing, which profoundly impact self-organization and transport in dense bacterial suspensions. Taken together, our work establishes fundamental knowledge for faithfully understanding the collective behavior of dense polar active fluids.

## Significance statement
Active fluids consisting of living cells or synthetic microswimmers display rich emergent behavior and nonequilibrium mechanical properties, which not only shed light on various biological processes across different length-scales, but also provide guidance to the design of autonomous fluidics and self-driven materials. The behavior of constituent self-driven units and their interaction with self-generated solvent flows underlie the emergent properties of active fluids. Here we studied the microscopic dynamics in dense 3D bacterial active fluids. Our work revealed an array of novel phenomena arising from activity-induced hydrodynamic effects, providing essential information for faithfully describing the emergent behavior of 3D bacterial active fluids. Our work clarifies how near-field hydrodynamics contribute to the collective behavior of dense active fluids.

## Introduction

Active fluids consist of self-driven units, such as living motile cells and synthetic microswimmers that are immersed in a liquid medium and engage in strong local interactions (1). Active fluids display rich self-organized collective behavior, including active turbulence (2-7), spontaneous directed flows (8-11), and large-scale spatiotemporal ordering (12, 13); they also exhibit unique nonequilibrium mechanical properties (14), such as superfluid-like viscosity reduction under external shear (15-18). These emergent collective behaviors and nonequilibrium mechanical properties not only shed light on various biological processes across different length-scales, such as cytoplasmic streaming (19) and collective epithelial migration (20), but also provide guidance to the design of autonomous fluidics for directed cargo transport (21-23), drug delivery (24), and microfluidic manipulation (25-27). The behavior of constituent self-driven units and their interaction with self-generated solvent flows underlie the emergent properties of active fluids. However, these relevant microscopic dynamics have been difficult to access in experiments, and mesoscale polarity ordering as a fundamental mechanism for the emergent dynamics was often taken *a priori* (2, 28-31).

Suspensions of swimming bacteria powered by rotating flagella have served as model systems of active fluids (2, 8, 9, 12, 15, 29, 32, 33). Here we developed a multichannel fluorescence imaging technique to study the microscopic dynamics in dense 3D active fluids consisting of flagellated bacteria. This technique allows simultaneous visualization of cell body, flagellum, and local fluid flows at high spatial resolutions. Unexpectedly, we discovered that the polarity of cells (i.e., the principal axis of active stress exerted by a solitary cell in a uniform flow) was randomly distributed in mesoscale flow regimes, and yet the resultant solvent flows displayed mesoscale order. Despite the apparent flow-polarity decoupling, the motion of cells relative to local solvent flows (in the Lagrangian reference frame) predominantly navigated upstream, with the self-advection speed overall enhanced by a flow-controlled constant. Numerical modeling with full hydrodynamic interactions reveals that the observed flow-polarity decoupling arises from the breakdown of the commonly held force-dipole assumption for anisotropic microswimmers: in the presence of near-field hydrodynamic interactions that depend on cell-cell pairwise configurations, the direction of total active forcing exerted by a swimming bacterium to the surrounding fluid no longer aligns with its polarity. The simulations suggest that near-field interactions serve as a new type of emergent, configuration-dependent active forcing, which profoundly impact self-organization and transport in dense bacterial suspensions. Taken together, our study establishes fundamental knowledge to understand the emergent behavior and nonequilibrium mechanics in dense polar active fluids. Meanwhile, the rich microscopic information provided by our findings may facilitate the modeling of dense active fluids via data-driven and machine-learning approaches (34, 35).

## Results

### Multichannel fluorescence imaging of microscopic dynamics in bacterial active fluids

We chose to study living active fluids consisting of the model flagellated bacterium *E. coli* (cell body length ~2 µm and width ~0.8 µm) and at a cell density of ~8 × $10^{10}$ cells/ml, which corresponds to a volume fraction of ~10.0% or an average cell-cell center distance ~2.3 µm. Bacterial active fluids in this regime exhibit mesoscale flow patterns with transient vortices and jets known as "active turbulence" (2, 29, 36) (Fig. S1a,b). To quantify the microscopic behavior of cells (including cellular advection and orientational dynamics) in response to these self-generated mesoscale flows, we need to simultaneously track cell bodies, flagella, and local fluid flows of the ambient solvent. The dense suspension of wildtype *E. coli* (strain HCB1737) was seeded with a small fraction of *E. coli* cells (strain YW263, with wildtype flagellar motility) whose body and flagella were fluorescently labeled for single-cell tracking; meanwhile, the suspension was supplemented with 0.27-µm fluorescent microspheres that served as fluid tracers (Methods). For this purpose we used an arbitrary-time-sharing multichannel fluorescence imaging scheme to arbitrarily set the exposure time of an sCMOS camera as well as the light illumination time of different wavelengths for cell body, flagellar filament and fluid tracer within one acquisition cycle (Fig. 1a; Methods). In combination with a suite of data processing algorithms for meticulous image alignment and object mapping, our method allows optimization of the signal-to-noise ratios of the three types of objects (Methods) to enable high-resolution and simultaneous measurement of the following microscopic quantities (two-dimensional projection on the focal plane; Fig. 1b,c, Video 1, Video 2): cell polarity $\boldsymbol{p}$ (i.e., flagellum-to-body direction, or the principal axis of active stress exerted by an isolated cell in a uniform flow), apparent single-cell velocity $\boldsymbol{u}_{\mathrm{cell}}$, and the velocity of local advective solvent flows $\boldsymbol{u}_{\mathrm{flow}}$ (which is the sum of flows collectively generated by nearby cells, including the tracked cell's own disturbance flow); the latter two were in the laboratory (Eulerian) reference frame. Furthermore, we defined the advection-corrected single-cell velocity or self-advection velocity $\boldsymbol{v}_{\mathrm{cell}}$ as $\boldsymbol{v}_{\mathrm{cell}} \equiv \boldsymbol{u}_{\mathrm{cell}} - \boldsymbol{u}_{\mathrm{flow}}$ (Fig. 1c, Video 2; Methods), which corresponds to the single-cell velocity relative to the measured local solvent flows (i.e., in the Lagrangian reference frame). A full list of microscopic quantities that were measured and analyzed experimentally can be found in Table S1 and Table S2 (Supplementary Materials). Our data analysis effectively sampled the trajectory of cells whose motion was primarily within the observation plane, and cells with a significant out-of-plane (i.e., vertical) velocity component were not taken into account (see Methods).

## Flow-polarity decoupling and upstream navigation in mesoscale flow regimes

With access to the microscopic quantities ($\boldsymbol{u}_{\mathrm{flow}}$; $\boldsymbol{u}_{\mathrm{cell}}$, $\boldsymbol{v}_{\mathrm{cell}}$, $\boldsymbol{p}$), we now sought to study the single-cell dynamics in self-generated mesoscale solvent flows by examining the relationship between solvent-flow velocity $\mathbf{u}_{\mathrm{flow}}$ and the quantities related to cell motion. To minimize the boundary effect, we focused on the mid-plane of the fluid chamber (~30 µm from all surfaces; Fig. 1a; Methods). While the apparent single-cell velocity $\mathbf{u}_{\mathrm{cell}}$ tended to align with $\mathbf{u}_{\mathrm{flow}}$, surprisingly, we observed that the advection-corrected single-cell velocity $\mathbf{v}_{\mathrm{cell}}$ predominantly pointed against the direction of $\mathbf{u}_{\mathrm{flow}}$ (~65% of all cases) (Fig. 2a). The probability distributions of the direction angles of $\mathbf{u}_{\mathrm{cell}}$ and $\mathbf{v}_{\mathrm{cell}}$ relative to $\mathbf{u}_{\mathrm{flow}}$ are mirror-symmetric with respect to each other (Fig. 2a). These results showed that the cells tended to navigate against the direction of local, mesoscale-ordered solvent flows generated by themselves, and the downstream bias of their apparent motion in the laboratory frame was due to local advection.

While bacteria can swim upstream against external flows near solid walls (37-39) or in the centerline of Poiseuille flows (40), the upstream navigation of cells against local self-generated solvent flows in dense active suspensions was not observed or predicted previously. This finding seemed to suggest that cell polarity tended to point against the self-generated solvent flows as well. However, our measurement revealed a surprising result: The angle between cell polarity $\mathbf{p}$ and the associated local solvent flow $\mathbf{u}_{\text{flow}}$ (denoted as $\theta$) was randomly distributed (Fig. 2b), with the $\mathbf{u}_{\text{flow}}$-$\mathbf{p}$ alignment parameter (defined as $\langle \mathbf{p} \cdot \hat{\mathbf{u}}_{\text{flow}} \rangle = \langle \cos\theta \rangle$, where $\hat{\mathbf{u}}_{\text{flow}}$ is the direction of $\mathbf{u}_{\text{flow}}$ and $\langle \cdot \rangle$ indicates the ensemble average of all cells in one experiment) being nearly zero. Furthermore, the in-plane Cartesian components of $\hat{\mathbf{u}}_{\text{flow}}$ and $\mathbf{p}$ did not show any correlation, i.e., $\langle \mathrm{p}_x \hat{\mathrm{u}}_{\text{flow},x} \rangle \approx 0$ and $\langle \mathrm{p}_y \hat{\mathrm{u}}_{\text{flow},y} \rangle \approx 0$ (Fig. S1c,d). As the vertical components of cell polarity $\mathrm{p}_z$ and local solvent flow direction $\hat{\mathrm{u}}_{\text{flow},z}$ are statistically indistinguishable from their in-plane Cartesian components at the mid-plane of the fluid chamber, it can be inferred that $\langle \mathrm{p}_z \hat{\mathrm{u}}_{\text{flow},z} \rangle \approx 0$ as well. Therefore, it is expected that the $\mathbf{u}_{\text{flow}}$-$\mathbf{p}$ alignment parameter is nearly zero (i.e, $\langle \mathbf{p} \cdot \hat{\mathbf{u}}_{\text{flow}} \rangle \approx 0$) even in the case of fully 3D measurement of $\mathbf{p}$ and $\mathbf{u}_{\text{flow}}$. In addition, cell polarity $\mathbf{p}$ was uncorrelated with the in-plane principal eigenvector of the flow field's strain rate tensor $E = (\nabla\mathbf{u}_{\text{flow}}{}^{\mathrm{T}} + \nabla\mathbf{u}_{\text{flow}})/2$ (Fig. S2), even though the two quantities were expected to correlate with each other in bacterial active fluids (41).

Taken together, we found that cell polarity $\mathbf{p}$ and the associated local solvent flow $\mathbf{u}_{\text{flow}}$ were decoupled. Since the solvent flow field in the dense bacterial active fluid exhibited mesoscale order (Fig. S1b; Methods), the absence of flow-polarity correlation at the mesoscale, or simply flow-polarity decoupling, suggested that the polarity is randomly distributed in mesoscale flow regimes, and that the microscopic active forcing is not determined by bacterial polarity alone as often assumed (see Discussion).

## Active hydrodynamics control the mesoscale order

It is intriguing how the mesoscale flow order and the collective upstream navigation of cells could have emerged despite the random distribution of local polarities. Naively, the advection-corrected single-cell velocity $\mathbf{v}_{\text{cell}}$ should align with cell polarity $\mathbf{p}$. Indeed, the directions of $\mathbf{v}_{\text{cell}}$ and $\mathbf{p}$ were correlated; however, the angle between them (denoted as $\theta_{\mathrm{v}}$) displayed a broad distribution, showing prominent misalignment (Fig. 2c). The misalignment between $\mathbf{v}_{\text{cell}}$ and $\mathbf{p}$ in the dense bacterial active fluid shows that cell-cell interactions (via activity-induced hydrodynamic flows and steric effects) play an important role in single-cell dynamics as well as in their collective transport. These interactions could also modulate the solvent flows, giving rise to the mesoscale flow order amid the random distribution of local polarities.

To delineate the contributions of active hydrodynamics and steric repulsion to the phenomena, we sought to tune the strength of active hydrodynamics while keeping steric effects constant. Following the approach in a recent study (36), we mixed motile and immotile cells to form a dense suspension with the total cell density fixed at ~$8 \times 10^{10}$ cells/ml (i.e., with the steric effects fixed); here the motile population consisted of wildtype *E. coli* cells (strain HCB1737) seeded with a small fraction of fluorescently labelled cells (strain YW263 with wildtype flagellar motility), and the immotile cells were derived from an *E. coli* strain with paralysed flagella (HCB84) (Methods). By adjusting the proportion of motile cells ($r_m$) that differ from the immotile ones only in motility (Fig. S3), the strength of hydrodynamic interactions (HI) can be tuned (Methods). First of all, the $\mathbf{u}_{\text{flow}}$-$\mathbf{p}$ alignment parameter $\langle \cos\theta \rangle$ remained $\sim 0$ as HI strength was increased (Fig. 3a, lower; Fig. S4, purple data points), i.e., the cell polarity was always random regardless of HI strength. By contrast, the extent of $\mathbf{v}_{\text{cell}}$-$\mathbf{p}$ misalignment (measured by a $\mathbf{v}_{\mathrm{cell}}$-$\mathbf{p}$ misalignment parameter $D_{\mathrm{v}} = 1 - \langle \cos\theta_{\mathrm{v}} \rangle$) quasi-linearly increased with the motile cell proportion $r_m$, from ~0.25 at the immotile limit $r_m \approx 0.001$ (Fig. S5) to ~0.7 at the motile limit $r_m \approx 1$ (Fig. 3a, upper; Fig. S4, black data points). Meanwhile, the mesoscale flow order (denoted as $M$; see Methods for definition) grew from near zero and reached a plateau as HI strength was increased (Fig. 3b, blue data points); similarly, the extent of upstream navigation (measured by an upstream navigation index defined as $-\langle \cos\varphi_{\mathrm{v}} \rangle$, where $\varphi_{\mathrm{v}}$ denotes the angle between $\mathbf{v}_{\mathrm{cell}}$ and the local solvent-flow velocity $\mathbf{u}_{\mathrm{flow}}$) grew and saturated concurrently with the mesoscale flow order (Fig. 3b, red data points; Fig. S4, orange data points). The two quantities ($M$ and $-\langle \cos\varphi_{\mathrm{v}} \rangle$) collapsed into a linear master curve for a wide range of motile cell ratios at several different cell densities (Fig. 3c). These results show that activity-induced hydrodynamic effects in the dense bacterial suspension control the mesoscale flow order and the upstream bias of $\mathbf{v}_{\mathrm{cell}}$.

**Active hydrodynamics result in overall mobility enhancement**

Having investigated the orientational dynamics in 3D dense bacterial active fluids, we next turned to the magnitude of cellular and solvent flow velocities, which are pertinent to the kinetic energy spectra and energy transfer in the active fluid (7, 42). The magnitudes of $\mathbf{u}_{\mathrm{flow}}$, $\mathbf{u}_{\mathrm{cell}}$ and $\mathbf{v}_{\mathrm{cell}}$ are denoted as $u_{\mathrm{flow}}$, $u_{\mathrm{cell}}$ and $v_{\mathrm{cell}}$, respectively. Surprisingly, despite the fact $\mathbf{u}_{\mathrm{cell}} \equiv \mathbf{u}_{\mathrm{flow}} + \mathbf{v}_{\mathrm{cell}}$, $u_{\mathrm{cell}}$ and $v_{\mathrm{cell}}$ display nearly identical distributions (Fig. 4a). As the HI strength or the motile-cell proportion $r_m$ was increased, the distributions of apparent single-cell speed $u_{\mathrm{cell}}$ and advection-corrected single-cell speed $v_{\mathrm{cell}}$ widened and expanded considerably into the high-speed region (Fig. 4a). The following experimental relation holds regardless of the strength of activity-induced hydrodynamic interactions (i.e., $r_m$), which will be further discussed in the next section:

$$\langle v_{\mathrm{cell}} \rangle \approx \langle u_{\mathrm{cell}} \rangle . \qquad \#(1)$$

More importantly, our data revealed two sets of simple linear relations. First, the average cellular and solvent-flow speeds $\langle v_{\mathrm{cell}} \rangle$, $\langle u_{\mathrm{cell}} \rangle$ and $\langle u_{\mathrm{flow}} \rangle$ all linearly increased with the HI strength $r_m$ (Fig. 4b). Second, the mean advection-corrected single-cell speed $\langle v_{\mathrm{cell}} \rangle$ and the mean solvent-flow speed $\langle u_{\mathrm{flow}} \rangle$ differed by a constant value ($\approx 20$ µm/s) independent of HI strength (Fig. 4b,c). This constant is close to $\langle v_{\mathrm{cell}} \rangle$ measured in the zero-HI limit (i.e., $r_m \approx 0.001$), which corresponds to the average self-advection speed of cells in the steric-interaction background of the dense bacterial suspension (denoted as $V_0$); we note that $V_0$ is ~30% lower than the intrinsic self-propulsion speed of isolated cells in the dilute limit $v_0$ (25.4 ± 10.6 µm/s; Fig. S3) due to steric hindrance. Therefore, we arrive at the following remarkable relation between the ensemble average of cellular speeds $\langle v_{\mathrm{cell}} \rangle$ and $\langle u_{\mathrm{flow}} \rangle$:

$$\langle v_{\mathrm{cell}} \rangle \approx V_0(1 + \kappa \langle u_{\mathrm{flow}} \rangle). \qquad \#(2)$$

Here $\kappa \approx$ (18.5 µm/s)$^{-1}$ is a constant obtained by fitting. While the frame rate of the our imaging setup does not allow direct measurement of flagellar rotation speed, the flagellar motor power output from single-cell motility (which determines the flagellar rotation speed) is not likely to be affected by hydrodynamic flows. Therefore we interpret Eq. 2 as the overall enhancement of single-cell mobility. This precise quantitative relation reveals that the magnitude of mobility enhancement is controlled by self-generated solvent flows, presumably via flow-induced suppression of single-cell wobbling; suppression of cell wobbling is known to enhance bacterial swimming speed (43, 44). Notably, $\langle v_{\mathrm{cell}} \rangle$ can be more than doubled as $r_m$ increased to 1.0 (Fig. 4b). This overall mobility enhancement is independent of the recently uncovered *local* mobility enhancement in dense bacterial suspensions that correlates with local polar order (13), although the two phenomena may share a similar mechanical origin. Alternatively, Eq. 2 can be interpreted as the reduction of apparent solvent viscosity. Here the apparent viscosity reduction applies to objects at the microscale (i.e., single cells). It is different from shear viscosity reduction measured at the mesoscale or macroscale in active suspensions under external shear (15-18); the latter was attributed to the onset of collective cellular motion that reduces the energy loss from viscous dissipation (18).

**Hydrodynamic anisotropy facilitates upstream navigation and impacts cellular speeds differentially**

Results presented in Fig. 3 and Fig. 4 show that activity-induced hydrodynamics in the dense active suspension account for the upstream navigation and lead to an overall enhancement of single cell mobility. Given the random distribution of cell polarity $\mathbf{p}$ relative to the local solvent-flow velocity $\mathbf{u}_{\mathrm{flow}}$ (Fig. 2b), we reasoned that the impact of HI on cell velocities (both direction and magnitude) may be anisotropic with regard to cell polarity. To examine this potential hydrodynamic anisotropy, we equally divided cells into three subpopulations with downstream polarity ($0° \leq \theta < 60°$), orthogonal polarity ($60° \leq \theta < 120°$), and upstream polarity ($120° \leq \theta \leq 180°$) (Fig. S6a). These three polarity subpopulations have a $\mathbf{u}_{\mathrm{flow}}$-$\mathbf{p}$ alignment parameter $\langle \cos\theta \rangle$ of ~0.9, ~0, and ~ -0.9, respectively (Fig. S6b).

Among the three polarity subpopulations, we found that cells with downstream polarity had the $\mathbf{v}_{\mathrm{cell}}$-$\mathbf{p}$ misalignment parameter $D_{\mathrm{v}} = 1 - \langle \cos\theta_{\mathrm{v}} \rangle$ growing most rapidly with the strength of hydrodynamic interactions (i.e., $r_m$) while holding the total cell number density fixed (Fig. 5a). That is, cells with downstream polarity were most strongly perturbed by HI; with increasing HI strength, they navigated more randomly and eventually lost all their navigation bias (Fig. 5b). By contrast, cells with upstream polarity were only weakly perturbed by HI (Fig. 5a) and retained their collective upstream navigation when HI strength was increased (Fig. 5b), whereas cells with orthogonal polarity were intermediately perturbed (Fig. 5a) and developed a weak upstream navigation bias (Fig. 5b) in the regime with saturated mesoscale flow order (Fig. 3b). These results revealed that the hydrodynamic anisotropy underlies the collective upstream navigation in dense bacterial active fluids.

Next, we investigated how the hydrodynamic anisotropy may influence the cellular speeds $u_{\mathrm{cell}}$ and $v_{\mathrm{cell}}$ for different polarity subpopulations. We found that, while the cellular speeds of all polarity subpopulations displayed a linear growth with the HI strength (i.e., $r_m$) similar to that in Fig. 4b, the subpopulation with downstream polarity had the fastest growing $\langle u_{\mathrm{cell}} \rangle$ (Fig. 5c) but the slowest growing $\langle v_{\mathrm{cell}} \rangle$ (Fig. 5d). Also the downstream-polarity subpopulation has downstream $\mathbf{u}_{\mathrm{cell}}$ bias (Fig. 5e); this result, together with the fastest growth of $\langle u_{\mathrm{cell}} \rangle$ with HI strength, suggests that local flow advection is dominant over self-propulsion for this subpopulation. Conversely, the upstream-polarity subpopulation had the fastest-growing $\langle v_{\mathrm{cell}} \rangle$ (Fig. 5d) but the slowest growing $\langle u_{\mathrm{cell}} \rangle$ (Fig. 5c); meanwhile their $\mathbf{u}_{\mathrm{cell}}$ directions became random as HI strength was increased (Fig. 5e). The converse trends in the relations between HI strength and cellular speeds (Fig. 5c,d) and cellular velocity directions (Fig. 5b,e) explain the equality $\langle u_{\text{cell}} \rangle \approx \langle v_{\text{cell}} \rangle$ in Eq. 1 as well as the mirror-symmetric distributions of cellular velocity directions shown in Fig. 2a.

## Discussion

Taken together, we have shown that activity-induced hydrodynamics control the collective transport in dense bacterial suspensions, giving rise to mesoscale order and overall mobility enhancement. Despite the mesoscale order in cellular and solvent transport, a surprising finding is that the polarity of cells was randomly distributed in mesoscale flow regimes. This flow-polarity decoupling suggests that the microscopic active forcing in dense bacterial suspensions is not determined by bacterial polarity alone, as required by the commonly used force-dipole model for isolated anisotropic microswimmers (45). In the force-dipole model, the head and tail of a microswimmer act as two point forces exerted to the surrounding fluid along the polarity (i.e., the vector pointing from tail to head) but in opposite directions, such that the principal axis of active stress $\sigma$ produced by the microswimmer aligns with its polarity $\boldsymbol{p}$ via the form $\sigma \sim \boldsymbol{pp}$ (45). This key assumption arises from two approximations; one is that the size of the bacterium can be treated as negligible, as it is generally much shorter than the characteristic length of changes in flow field in the suspension, and the other is that near-field hydrodynamic interactions (which depend on cell-cell pairwise configurations or relative distance between cells and can be described by lubrication theory) are negligible.

However, the characteristic length of the flow field in dense bacterial suspensions is not much longer than the full length of one typical bacterium (flagella and cell body summing up to ~7-10 µm), and in our experiments bacteria are frequently close to each other (average cell center distance ~2.3 µm), where near-field hydrodynamics have been proven to be significant (46, 47). As demonstrated in Fig. S7, the active forces exerted by bacterial body and flagella to the surrounding fluid can be significantly misaligned with cell polarity $\boldsymbol{p}$, either for a solitary bacterium swimming in a non-uniform flow (Fig. S7a) or for two bacteria close to each other in otherwise quiescent fluid (Fig. S7b). Therefore, the force-dipole assumption would not be valid in dense bacterial suspensions.

To examine the consequence of the breakdown of force-dipole assumption in dense bacterial suspensions, we performed full hydrodynamic modeling of 3D bacterial suspensions that incorporates an analytic description of the hydrodynamic interactions between cells in both the far-field limit and the near-field limit (Methods). In the model, each bacterium is simplified as two spheres with no-slip boundary connected by a rigid rod (Fig. 6a): one sphere for the bacterial body and the other for the flagellar bundle, where the flagellar sphere is propelled by a phantom force $\boldsymbol{F}_{act}$ (with constant magnitude pointing from the center of flagellar sphere to the center of body sphere) provided by the self-spinning of the flagellar not treated explicitly (47, 48). This two-sphere composite captures the natural anisotropic geometry of each *E. coli* bacterium. By defining the intrinsic self-propulsion velocity $\boldsymbol{u}_0 = \boldsymbol{F}_{act}/(6\pi\mu R_t)$ for each bacterium where $\mu$ and $R_t$ are viscosity of water and size of flagellar sphere, respectively, the dynamics of the system in the overdamped limit is described by:

$$(\mathbf{I} + \mathbf{M}_\infty \cdot \boldsymbol{\xi}_{\text{near}}) \cdot (\boldsymbol{U} - k_B T \nabla \cdot \boldsymbol{\xi}^{-1}) = \mathbf{M}_\infty \cdot \left(\boldsymbol{F}^{\text{P}} + \boldsymbol{F}^{\text{B}}\right). \qquad \#(3)$$

Here the resistance tensor $\boldsymbol{\xi} = \mathbf{M}_\infty^{-1} + \boldsymbol{\xi}_{\text{near}}$ fully describe the hydrodynamic interactions for any given cellular configuration, by accounting for both the far-field hydrodynamics through mobility tensor $\mathbf{M}_\infty$ and the near-field hydrodynamics through lubrication $\boldsymbol{\xi}_{\text{near}}$; with both tensors analytically available as described in the classical Stokesian Dynamics Simulation method (49, 50). For the rest of the terms, $k_\text{B}T\nabla \cdot \boldsymbol{\xi}^{-1}$ is the so-called Brownian drift, a deterministic part of the solvent thermodynamic effect that becomes significant when bacteria are close to each other or close to the no-slip boundaries (48); $\boldsymbol{U} \equiv (\boldsymbol{U}_1, \ldots, \boldsymbol{U}_i, \ldots, \boldsymbol{U}_N)^T$ with $\boldsymbol{U}_i \equiv \left(\boldsymbol{u}_{i\text{b}}, \boldsymbol{\omega}_{i\text{b}}, \boldsymbol{u}_{i\text{t}} - \boldsymbol{u}_i^0, \boldsymbol{\omega}_{i\text{t}}\right)^T$ being the translational/rotational velocity vector characterizing the motions of the body-sphere and tail-sphere, respectively, for the $i$-th bacterium; $\boldsymbol{F}^\text{P} \equiv \left(\boldsymbol{F}_1^\text{P}, \ldots, \boldsymbol{F}_i^\text{P}, \ldots, \boldsymbol{F}_N^\text{P}\right)^T$ with $\boldsymbol{F}_i^\text{P} \equiv (\boldsymbol{F}_{i\text{b}}, \boldsymbol{L}_{i\text{b}}, \boldsymbol{F}_{i\text{t}}, \boldsymbol{L}_{i\text{t}})^T$ being the intrinsic, nonhydrodynamic forces/torques for the body-sphere and tail-sphere, respectively, for the $i$-th bacterium; and $\boldsymbol{F}^\text{B}$ represents the stochastic forces/torques (thermal effect of the solvent). In the simulation we chose to neglect this random noise part by setting $\boldsymbol{F}^\text{B} = 0$.

The dynamics of the system was simulated by solving the motion $\boldsymbol{U}$, and force/torque $\boldsymbol{F}^\text{P}$ for all bacteria according to Eq. 3 (Methods). The simulated system displayed collective motion with the kinetic energy spectrum following a $\sim k^{-3}$ scaling behavior similar to that in the experiment (Fig. S1f)(51). Due to limitations of computational efficiency, the cell number density in the simulations was chosen as ~0.8× $10^{10}$ cells/ml (i.e., average cell center distance ~5 µm), which is one order of magnitude smaller than that used in the experiments. For this reason, cells in this simulation do not always have near neighbors as their experimental counterparts; so we analyzed the dynamics of cells with and without near neighbors, respectively. For cells with near neighbors (i.e., those having at least one neighbor with surface distance less than 2% of the body radius), our simulation results as shown in Fig. 6b-e are in good agreement with the experimental results in Fig. 2. Importantly, the model reproduces the random distribution of single-cell polarity relative to the local solvent flows (Fig. 6d versus Fig. 2b); also the modelled cells display upstream navigation against the local solvent flows (Fig. 6c versus Fig. 2a). By contrast, for cells without near neighbors, the polarity and navigation direction both tend to align with the local solvent flows (Fig. 6f-i). We note that our model does not account for the elastohydrodynamic coupling between flagellar hook mechanics and the external flows that gives rise to the flow-induced modification of cell wobbling. Therefore it is beyond our current modeling work to examine our hypothesis that the observed overall mobility enhancement arises from flow-induced suppression of cell wobbling.

These simulation results clearly demonstrate that the flow-polarity decoupling and upstream navigation are both rooted in near-field interactions between cells. The precise role of near-field interactions can be understood conceptually as follows. Based on Eq. 3, the solvent flow field generated by individual cells (denoted as $\mathcal{U}_\text{flow}$) is no longer only related to the intrinsic active forcing of individual cells $\boldsymbol{F}^\text{P} \equiv \left(\boldsymbol{F}_1^\text{P}, \ldots, \boldsymbol{F}_i^\text{P}, \ldots, \boldsymbol{F}_N^\text{P}\right)^T$, but instead $\boldsymbol{F}^\text{P} - \boldsymbol{\xi}_\text{near} \cdot (\boldsymbol{U} - k_\text{B}T\nabla \cdot \boldsymbol{\xi}^{-1})$ (see Methods). This formula reveals a new type of active forcing in the form of $\boldsymbol{\xi}_\text{near} \cdot (\boldsymbol{U} - k_\text{B}T\nabla \cdot \boldsymbol{\xi}^{-1})$**,** which arises from the near-field hydrodynamic interactions characterized by the resistance matrix $\boldsymbol{\xi}_\text{near}$ and thus depends on cell-cell pair configurations**.** As a result, $\mathcal{U}_\text{flow}$ has two contributions, one from the intrinsic or self-part of active forcing $\sim\boldsymbol{F}^\text{P}$ (which is correlated with but not perfectly aligned with an individual cell's polarity), and the other from the coupling part of active forcing $\sim\boldsymbol{\xi}_\text{near} \cdot (\boldsymbol{U} - k_\text{B}T\nabla \cdot \boldsymbol{\xi}^{-1})$ related to cell-cell pairwise configurations. The self-part of $\mathcal{U}_\text{flow}$ is still correlated with cell polarity, and its local average is ~ 0 due to locally random polarity distribution. Therefore, the coupling part of $\mathcal{U}_{\mathrm{flow}}$ arising from near-field hydrodynamic interactions must be the primary source of mesoscale-ordered solvent flows. At the single-cell scale, the configuration-dependent active forcing can act differentially across the cell body and the flagella, providing mechanical torque to randomize the cell polarities and causing flow-polarity decoupling. Moreover, as suggested by the anisotropy of hydrodynamic effect on single-cell velocity (Fig. 5a,b), the perturbation by the configuration-dependent active forcing may depend on the polarity of a cell due to the fore-aft asymmetry of bacterial cells, impacting upstream-orienting cells less than others and resulting in the observed collective upstream navigation.

Our discovery of flow-polarity decoupling in pusher suspensions provides a fresh perspective on the emergent mechanics of active turbulence. Large-scale simulations of dense microswimmer suspensions with both near-field and far-field hydrodynamic interactions—such as confined spheroidal squirmers in a quasi-2D slit geometry (52) and unconfined spherical squirmers in 3D (53)—have demonstrated active turbulence. However, previous simulations of dense spheroidal pusher suspensions did not resolve the ambient solvent flow field, leaving the flow-polarity relation unmeasured. Furthermore, while prior numerical studies using spherical squirmer models observed orientation-velocity decoupling primarily in puller suspensions (e.g., (53)), dense pusher squirmers typically require explicit, artificial neighbor-alignment rules for 3D active turbulence to emerge. This discrepancy highlights a fundamental limitation of treating elongated microswimmers as spheres with prescribed surface slip velocities: the rotation of a spherical squirmer couples exclusively to local vorticity and is incapable of coupling to the local strain rate tensor. In contrast, our two-sphere model explicitly represents the anisotropic, multi-part geometry of the flagellated bacteria used in the current experimental study, where the differing flow velocities experienced across the body and the tail naturally reproduce the coupling between cell orientation and the local strain rate. The configuration-dependent near-field lubrication forces $\boldsymbol{\xi}_{\mathrm{near}} \cdot (\boldsymbol{U} - k_{\mathrm{B}} T \nabla \cdot \boldsymbol{\xi}^{-1})$ can act differentially across the body-sphere and the tail-sphere, providing the mechanical torque for flow-polarity decoupling, while giving rise to mesoscale active turbulence without artificial neighbor alignment. Currently our computational resources have limited the cell density in simulations, and future improvement of the algorithm will be needed to simulate denser suspensions with fully developed mesoscopic active turbulence.

## Conclusion

To summarize, we have studied the microscopic dynamics in 3D dense bacterial active fluids and revealed an array of novel phenomena arising from activity-induced hydrodynamic effects. In particular, combining experiment and hydrodynamic modeling, we found that near-field hydrodynamic interactions cause flow-polarity decoupling in the dense active fluid via a type of emergent, configuration-dependent active forcing. Our work establishes fundamental knowledge of microscopic dynamics in dense polar active fluids, providing essential information for faithfully describing the emergent collective behavior of 3D polar active fluids consisting of living organisms or synthetic microswimmers.

Our findings highlight the indispensable role of near-field hydrodynamics in the collective behavior of dense active fluids. Although this essential role was recognized (46, 47), near-field hydrodynamic interactions have been difficult to probe in either experiments or theories. The rich microscopic details offered by our findings and the large-scale simulations of the hydrodynamic interactions for dense active fluids utilized here can inform data-driven and machine-learning approaches (34, 35) for developing active fluid models that accurately account for near-field hydrodynamics, which may in turn help to further understand how near-field hydrodynamic interactions modulate the solvent and cellular transport reported here. Alternatively, the direction of advection-corrected single-cell velocity can be taken as an effective polarity tightly coupled to the self-generated mesoscale solvent flows, which accounts for both the intrinsic and coupling part of microscopic active forcing at the single-cell level. Modified with this effective polarity, current modeling framework for active fluids incorporating the statistical knowledge of near-field hydrodynamics may efficiently describe the dynamics of dense microswimmer suspensions.

## Methods

### Bacterial strains and growth

We used three *E. coli* strains in this study (Fig. S3): (i) strain HCB1737 with wild-type motility (denoted as 'Wildtype'), which carries a single amino acid (cysteine) substitution in flagellin protein FliC (in the background of *E. coli* AW405) that can react to thiol-reactive fluorescent dyes (54), hence the flagellar filaments of this strain can be fluorescently labeled by maleimide dyes (from Howard C. Berg, Harvard University); (ii) YW263 (*E. coli* HCB1737 transformed with the plasmid pKatushka2S-B [FP763; Evrogen]) carrying Ampicillin resistance and expressing red fluorescent protein Katushka2S constitutively, i.e., the cell body is fluorescently labeled; (iii) HCB84, an immotile strain with paralysed flagella (*E. coli* AW405 Δ*motA*, from Howard C. Berg, Harvard University).

All three bacterial strains followed the same growth protocol. Single-colony isolates were first inoculated to 2.5 mL LB medium (1% Bacto tryptone, 0.5% yeast extract, 0.5% NaCl; wt/vol) and grown as a day culture to exponential phase (~4-9 hr, $OD_{600} < 2.0$) in 12 mL plastic culture tubes (with shaking at 180 rpm and incubated at 30°C). Then the day culture was diluted to $10^5$ cells/mL in 25 mL LB medium and grown as an overnight culture for 14 hr (in a 250 mL Erlenmeyer flask with shaking at 180 rpm and incubated at 30°C). The overnight culture was stored inside a refrigerator (4 °C) to preserve cell motility until use. The duration of storage at 4 °C should be less than 6 hr. Ampicillin (100 µg/mL) was added into the LB medium for the growth of the strain YW263 to maintain the plasmid. All three strains had the same cell length in the collected overnight culture (Fig. S3). Strains HCB1737 and YW263 had the same swimming speed (Fig. S3).

### Sample preparation and flagella labeling

To prepare a bacterial suspension, several mL overnight cultures of motile HCB1737 and immotile HCB84 prepared above and stored at 4 °C were mixed at desired cell density ratios (to obtain desired values of motile cell proportion $r_m$; cell density was measured by $OD_{600}$ prior to mixing) and then treated with 5 U/mL DNase I (Invitrogen™, catalog number: 18047019) and 0.05 mg/mL RNase A (Invitrogen™, catalog number: EN0531) for ~20 minutes to remove extracellular polymers released by cells in the overnight culture; note that LB medium in the overnight cultures contains sufficient $Mg^{2+}$ ions for enzymatic activities. Next, flagellum-stained YW263 cells (see below for the procedures of flagella staining) were seeded into the mixed and treated bacterial culture at a cell number fraction of ~0.1% (for dense active suspensions); the number fraction was chosen such that the flagellum-stained cells in the field of view during image acquisition were of a sufficiently high density but they rarely overlapped in an image frame. To remove the unreacted dyes, the mixed bacterial suspension was then centrifuged for 3 min with 7000×g in a benchtop centrifuge (Eppendorf™ Model 5418 Microcentrifuge). The supernatant was removed, and the pellet was re-suspended gently with ~1 mL motility buffer (10 mM potassium phosphate, pH 7.0, 67 mM NaCl, 0.1 mM EDTA, and 0.001% vol/vol Tween 20). Then the resuspended suspension was washed again following the same procedures, except that the pellet was now re-suspended with ~10 µL nuclease-supplemented motility buffer (i.e., motility buffer supplemented with 100 U/mL DNase I, 1 mg/mL RNase A, and 1 mM $MgSO_4$). Finally, the suspension was adjusted to desired cell densities (measured by OD600) with nuclease-supplemented motility buffer together with suspensions of fluorescent microspheres (0.27 μm in diameter suspended in DI water, added at a final concentration of 0.01% wt/vol; Thermo Scientific Fluoro-Max dyed green aqueous fluorescent particles, cat. no. G300). Note that the 0.27 μm microspheres had a Stokes number of ~$10^{-9}$ (computed as $\sim t_0 u_0 / l_0$, where $t_0$, $u_0$, and $l_0$ were characteristic scales of particle relaxation time, flow speed, and mesoscale vortex size in the flow system), so they could faithfully trace the streamlines in the solvent flow.

To stain bacterial flagella, we used Alexa Fluor™ 647 C2 Maleimide (dissolved in DMSO and stored as a 5 mg/mL stock solution; Invitrogen™, catalog number: A20347). 100 μL of the YW263 overnight culture prepared and stored at 4 °C as described above was transferred to a 12 mL plastic culture tube and diluted in LB medium by 3-fold and supplemented with Alexa Fluor™ 647 dye solution to a final dye concentration of 0.1 mg/mL. The cell and dye mixture were shaken for 20 minutes (30 °C and 180 r.p.m., in a 12 mL plastic culture tube, under a dark condition) for reaction. As soon as the dye reaction was completed, the cell and dye mixture were added to the nuclease-treated bacterial suspensions made from overnight cultures (see above). The flagellum-stained YW263 cells had the same cell body length and swimming speed compared to wild-type HCB1737 and unstained YW263 cells under identical observation conditions (Fig. S3). This ensured that the single-cell dynamics of flagellum-stained YW263 cells were representative of the unstained wildtype cells.

**Microfluidic experiments**

The microfluidic devices (PDMS-glass chambers) used in the study were prepared as follows. We used standard soft photolithography to fabricate the desired SU-8 (Kayaku Advanced Materials; SU-8 2000 series) patterns spincoated on silicon wafers as molds, following the manufacturer's protocols and using low-precision film masks. Polydimethylsiloxane (PDMS) (Dow Corning SYLGARD™ 184 silicone elastomer, base to curing agent = 10:1) was cast on the SU-8 mold and placed in a vacuum to release bubbles. Then the PDMS was cured on the SU-8 mold overnight (>12 hr) at 50 °C in an oven. When the cured PDMS was cooled to room temperature, we peeled it off and punched holes on the inlet and outlet wells with a ~1 mm diameter hole puncher. The PDMS sheet was cleaned by supersonic washing inside an isopropanol bath for ~30 min, dried at 50 °C for ~30 min to remove the residue isopropanol, and then attached to a cleaned glass slide when cooled to room temperature, forming a PDMS-glass device; the glass slide was supersonically cleaned in an ethanol bath with saturated sodium hydroxide (~30 min), rinsed with DI water for three times, blown off residue water droplets, dried under 50 °C for ~30 min, and stored at room temperature until further use. Finally, the PDMS-glass device was heated at 50 °C overnight while being pressed by a heavy weight in order to strengthen the bonding.

For the study of single-cell dynamics in self-generated flows in dense bacterial suspensions, we injected the bacterial suspension prepared above (seeded with flagellum-stained YW263 cells and fluorescence microspheres) into a circular PDMS-glass chamber (~2 mm in diameter and ~60 μm in thickness) via the inlet hole on the inlet well by pipetting. The dense active suspension filled up the chamber while the air inside the chamber was drained via the outlet channel. The inlet and outlet channels were narrow (~ 50 μm in width and ~60 μm in height) and the residue flow from the inlet to the outlet was negligible. Once the suspension injection was completed, we waited for ~1 minute until the dense active suspension was stabilized and then acquired images within 10 minutes from the injection, in order to avoid loss of bacterial motility or cluster formation.

**Arbitrary-time-sharing multichannel fluorescence microscopy**

To perform the arbitrary-time-sharing multichannel fluorescence microscopy, we built the optical path on an inverted microscope (Nikon Eclipse Ti2-E). As illustrated in Fig. 1a, the excitation light was provided by a multi-channel LED light source (Lumencor Spectra III light engine). Three LED channels were used with fixed output powers, namely the cyan channel (~520 mW output power), the green channel (480-490 mW output power), and the red channel (450-500 mW output power), for imaging microspheres, cell bodies and flagella, respectively. A triple-band filter-set was used; it was composed of an emission filter (Semrock 515/588/700 nm BrightLine® triple-band bandpass filter), a dichroic mirror (Semrock 488/543/635 nm BrightLine® triple-edge dichroic beamsplitter), an excitation filter (Semrock 475/543/702 nm BrightLine® triple-band bandpass filter). An additional notch filter (Chroma ZET561TopNotch™ 561nm Notch Filter) was placed in front of the excitation filter at the backport of the microscope to prevent breaching of green light. No special optical settings or attachments were applied to the rest of the microscope's optical path. We used an sCMOS camera (Photometrics Prime BSI Express) to acquire fluorescent images.

To acquire fluorescent images, we set the sCMOS camera to the level trigger-in mode and the rolling shutter expose-out mode. The camera received a trigger-in signal to expose and generated an almost identical expose-out signal except for a ~2.5 ms expose-out delay (see Fig. S8a, within the duration marked with black) for 2048 pixel × 512 pixel ROI (in the 11-bit sensitivity gain mode). The electronic trigger-in signal was provided by a signal generator (FeelTech FY6300-20M; illustrated in Fig. 1a) that can produce arbitrary waveforms in a programmable manner. The expose-out signal was divided by a custom-made signal splitter programmed with an Arduino microprocessor chip to trigger and control the three LED channels on the LED light source. In our camera mode setting, each rising and falling edge of the trigger-in signal initiated and ended the hardware events (i.e., camera exposure and light illumination) associated with the acquisition of each frame, respectively.

With the electronic triggering system, we were able to capture an arbitrary sequence of image frames, and each frame could be exposed by any LED channel with an arbitrary exposure time. In our experiments, we defined an acquisition cycle (lasting for 50 ms) as the following consecutive row of 5 frames (see Fig. 1b and S7a), with the corresponding duration of the trigger-in signal and the LED channel for fluorescence excitation specified in brackets: the flagellum frame (trigger-in signal lasting 19.5 ms plus 0.5 ms interval; excited by red LED), the first cell-body frame (9.5 ms plus 0.5 ms interval; green LED), the first microsphere frame (4.5 ms plus 0.5 ms interval; cyan LED), the second cell-body frame (9.5 ms plus 0.5 ms interval; green LED), and the second microsphere frame (4.5 ms plus 0.5 ms interval; cyan LED). With the ~2.5 ms expose-out delay subtracted, the actual exposure times for the 5 image frames were as follows: ~17 ms, ~7 ms, ~2 ms, ~7 ms, and ~2 ms (see Fig. 1b and S7a, where the exposure durations were marked with colored bars). In this setting of the acquisition cycle, we achieved an effective frame rate of 20 fps for flagellum frames, of 40 fps for cell-body frames, and of 40 fps for microsphere frames (Fig. 1b).

Images were acquired using a 40x objective (Nikon CFI Plan Fluor, NA 0.75, depth of focus ~1 μm). The corresponding ROI was 332.8 μm × 83.2 μm. The observation was made at the mid-plane of the microfluidic devices far from the lateral boundaries of the chamber confinement, so the lateral boundary effect was negligible. A typical recording for the study of single-cell dynamics in self-generated flows in bacterial active fluids lasted for ~30 s (~3000 total frames). We next extracted the solvent flow field, flagellar position and shape, and cell body's position and orientation, based on the captured images.

## Image processing and data analysis

**Python modules**. We used the following Python modules for image processing, data analysis, and data presentation: NumPy module (numpy 1.26.2); OpenPIV module (openpiv-python v0.25.0); OpenCV module (opencv-python 4.7.0.72); SciPy module (scipy 1.11.4); scikit-image module (scikit-image 0.20.0); Matplotlib module (matplotlib 3.8.0).

**Solvent-flow field calculation**. We used the OpenPIV Python module to calculate the solvent-flow field velocity vectors by comparing consecutive pairs of microsphere fluorescence image frames. We used a search window of 15.6 μm with an overlap size of 7.8 μm. Three filters from the openpiv.validation module were used for data validation: the signal-to-noise filter (.sig2noise_val, threshold = 1.0), the global standard deviation filter (.global_std, threshold = 3 times of standard deviation of the flow field), and the local median validation (.local_median_val, threshold = 100 μm/s, size = 3) filter. After each filter, invalid vectors were replaced using an iterative image inpainting algorithm in the openpiv.filters module (.replace_outliers, default parameters).

The PIV-calculated solvent-flow field vectors contained both advective solvent flow and Brownian noise of microspheres. To reduce the Brownian noise, we further smoothed the PIV-calculated flow field vectors to obtain the noise-reduced solvent-flow field $\mathbf{u}_{\text{flow}}$ within an acquisition cycle (from 0 ms to 50 ms; denoted as $\mathbf{u}_{\text{flow}}^{0\to 50\text{ ms}}$) by averaging the five temporally adjacent PIV-flow field vectors (Fig. S8b):

$$\mathbf{u}_{\text{flow}} \equiv \mathbf{u}_{\text{flow}}^{0\to 50\text{ ms}} \approx (\mathbf{u}_{\text{flow}}^{-52.5\to -17.5\text{ ms}} + \mathbf{u}_{\text{flow}}^{-17.5\to -2.5\text{ ms}} + \mathbf{u}_{\text{flow}}^{--2.5\to 32.5\text{ ms}} + \mathbf{u}_{\text{flow}}^{32.5\to 47.5\text{ ms}} + \mathbf{u}_{\text{flow}}^{47.5\to 82.5\text{ ms}})/5.$$

Here $\mathbf{u}_{\text{flow}}^{\text{t1}\to\text{t2}}$ was the PIV-calculated flow field based on microsphere image frames acquired at t1 and t2 (mid-frame time points). Note that the 2.5 ms exposure-out delay shifted the mid-frame time points in all frames with a constant, so it does not affect the velocity field calculation. This smoothing is valid, as the acquisition cycle duration of 50 ms and the smoothing window of 135 ms were much shorter than the correlation time of bacterial collective motion in dense active suspensions (typically on the order of a few seconds) and they were comparable to the characteristic free-swimming time interval between cell-cell collisions in a dense bacterial suspension. With this smoothing, the mean solvent-flow speed $\langle u_{\text{flow}} \rangle$ in suspensions consisting of entirely immotile cells at the same number density was close to zero (~0.23±0.16 μm/s, mean±s.d., N≈ 66,000 event counts). A flow speed filter was applied to discard events with flow speed <0.5 μm/s (approximately mean+1.5 s.d. of noise level), ensuring that all single-cell events used for the data analysis had well-defined local flow velocities.

We adopted the method in Ref. (36, 55) to quantify the spatial order of the solvent-flow field, The local velocity coherency $c$ for a flow vector $\mathbf{u}_{\text{flow}}(x,y)$ on $(x,y)$ was defined by comparing it with its $N$ adjacent vectors $\mathbf{u}_{\text{flow}}(x',y')$:

$$c=\frac{1}{N}\sum_{(x\prime,y\prime)\in R}\frac{\mathbf{u}_{\text{flow}}(x',y')\cdot\mathbf{u}_{\text{flow}}(x,y)}{u_{\text{flow}}(x',y')u_{\text{flow}}(x,y)}.$$

Here $u_{\text{flow}}$ is the magnitude of $\mathbf{u}_{\text{flow}}$ and the $N$ adjacent vectors are within area $R$ centered at $(x,y)$; the area was chosen as 31.2 μm × 31.2 μm (5×5 vectors). The mesoscale flow order $M$ was then defined as the fraction of vectors with local velocity coherency $c>0.5$:

$$M=\frac{N_{c>0.5}}{N_{\text{all vectors}}}.$$

Here $N_{\text{all vectors}}$ refers to the number of vectors in all frames in a time-lapse recording. $M$ equals 1 if all flow velocity vectors are perfectly parallel to each other, and nearly 0 if the flow field is random.

**Recognition of cell body and flagellum**. We took two steps to recognize cell body and flagellum, an image binarization step and a labeling step. For image binarization, from the original cell body and flagellum images (Fig. S9a), we first used the Wiener filter (scipy.signal.wiener) for denoising and the rolling ball algorithm (cv2.morphologyEx with morphological opening, rolling ball size = 8.125 μm) for background normalization (Fig. S9b). Next, the Canny edge detector (skimage.feature.canny, sigma = 2.0, low threshold = 1.0, high threshold for cell body = 20, high threshold for flagellum = 50) was used to detect cell body and flagellum edges in the background-normalized images (Fig. S9c). We then connected detected edges into regions of patterns with morphological closing (Fig. S9d, skimage.morphology.dilation followed by skimage.morphology.erosion with a flat and disk-shaped footprint, disk radius for cell body = 3, disk radius for flagellum = 6), resulting in a binarized image with regions of patterns corresponding to cell bodies or flagellar bundles. Our binarization algorithm had better performance compared to a simple binary thresholding algorithm when recognizing cell body and flagellum patterns with different fluorescence intensities. In the next step, the connected regions of patterns in binarized images were labeled (skimage.morphology.label, 2-connectivity) as single-cell bodies and flagellar bundles (Fig. S9e). Unreasonably large regions were filtered out as dirt spots or as connected patterns of multiple cell bodies or flagellar bundles.

With the single-cell body and flagellum regions labeled, we then measured the region properties (skimage.measure.regionprops). The center position of a cell body or a flagellar bundle was defined as the geometric centroid of the recognized region. In the meantime, by calculating the normalized second central moment matrix of the recognized region and comparing them to an ellipse with the same second central moment matrix, we obtained the long and short axes of the region, which were equal to the major and minor axes of the ellipse, respectively (Fig. S9f, g). The obtained cell body long axis is denoted as $b\mathbf{n}_{\text{cell}}$ with orientation $\mathbf{n}_{\text{cell}}$, while the flagellum long axis is denoted as $f_1\mathbf{n}_{\text{fla}}$ with orientation $\mathbf{n}_{\text{fla}}$. For flagellar bundles, Fig. S9h shows the recognized center position (blue dot), vertex position (cyan dot, which is the end of the half-short axis), and tail-end position (yellow dot, which is the end of the half-long axis).

**Mapping of cell body and flagellum**. We obtained the flagellum and cell body center positions in the flagellum and cell body frames, respectively, so we could map or link these positions acquired at different frames for further analysis (Fig. S9i-n). For the cell body center positions in the current cell body frame (at 25 ms, the first cell-body frame of an acquisition cycle; Fig. S9j), we mapped them to the cell body center positions in the previous cell body frame (at -10 ms, the second cell-body frame of the previous acquisition cycle; Fig. S9i), to the cell body center positions in the next cell body frame (at 40 ms, the second cell-body frame of the current acquisition cycle; Fig. S9k), and to the corresponding flagellum center positions (at 10 ms, the flagellum frame of the current acquisition cycle; Fig. S9l), using the same mapping algorithm detailed below.

In the mapping algorithm, Voronoi polygons were generated based on the cell body center positions in the current cell body frame (scipy.spatial.Voronoi). For a cell body or flagellar bundle position in another frame, if it was the only one in a Voronoi polygon associated with a specific current cell body center position, it was mapped to that current cell body center position. If there were more than one feature positions in the other frame that could be found within a specific Voronoi polygon, we aborted the mapping to avoid ambiguity. On the other hand, for a cell body center position in the current cell body frame, if no feature position in the other frame could be found within the Voronoi polygon associated with it or if its Voronoi polygon failed to be generated, we simply mapped it to the nearest feature position in the other frame. Finally, we filtered out mappings between two positions with unreasonably long distances. Fig. S9m shows the mapping of cell body positions in three adjacent frames and the corresponding displacements are displayed. Meanwhile, Fig. S9n shows the mapping between the body center and flagellum center positions.

**Calculation of single-cell velocities and polarity**. The apparent cell velocity $\mathbf{u}_{\text{cell}}$ in the laboratory reference frame in an acquisition cycle was computed based on the cell body center positions $(x, y)$ in the mid-time point of three consecutive cell body frames (Fig. S8c):

$$\mathbf{u}_{\text{cell}} \equiv \mathbf{u}_{\text{cell}}^{0\to 50\text{ ms}} \approx \frac{1}{2}\left(\mathbf{u}_{\text{cell}}^{-10\to 25\text{ ms}} + \mathbf{u}_{\text{cell}}^{25\to 40\text{ ms}}\right) =$$
$$\frac{1}{2}\left[\frac{(x,y)^{\text{at }25\text{ ms}} - (x,y)^{\text{at }-10\text{ ms}}}{35\text{ ms}} + \frac{(x,y)^{\text{at }40\text{ ms}} - (x,y)^{\text{at }25\text{ ms}}}{15\text{ ms}}\right].$$

The rationale of this averaging is that a cell could be regarded as moving in a ballistic manner within one acquisition cycle of 50 ms. Again the 2.5 ms exposure-out delay shifted the mid-time points in all frames with a constant, so it does not affect the velocity calculation. By subtracting the local solvent-flow velocity $\mathbf{u}_{\text{flow}}$ at the position of a specific cell, we deduced the advection-corrected velocity $\mathbf{v}_{\text{cell}}$ of the cell as $\mathbf{v}_{\text{cell}} = \mathbf{u}_{\text{cell}} - \mathbf{u}_{\text{flow}}$; here the local solvent-flow velocity $\mathbf{u}_{\text{flow}}$ at the cell's position was linearly interpolated (scipy.interpolate.RegularGridInterpolator) using the grid-meshed PVI-calculated mesoscale advection flow field with noise reduction as described in the section "Solvent-flow field calculation".

To calculate cell polarity $\mathbf{p}$, we first determined the cell body center position corresponding to the mid-time point of the flagellum frame, which is denoted as $(x,y)_{\text{flagellum frame}}^{\text{at }10\text{ ms}}$. This position was obtained by interpolation based on the cell body position at the mid-time point of the first cell body frame $(x,y)^{\text{at }25\text{ ms}}$ and the apparent cell velocity $\mathbf{u}_{\text{cell}}$:

$$(x,y)_{\text{flagellum frame}}^{at\ 10\text{ ms}} = (x,y)^{\text{at }25\text{ ms}} - \mathbf{u}_{\text{cell}} \times 15\text{ ms}.$$

After connecting the flagellum center position to $(x, y)^{at\ 10\ \mathrm{ms}}_{\mathrm{flagellum\ frame}}$, we obtained the direction vector from flagellum center to cell body center $a\mathbf{p}$ (with unit vector $\mathbf{p}$) (Fig. S9h, n). The unit vector $\mathbf{p}$ was taken as the cell polarity. Alternatively, we can define a direction vector from flagellum vertex to cell body center $a_1\mathbf{p}_1$ or from the flagellum tail to flagellum vertex $a_2\mathbf{p}_2$, and the unit vector $\mathbf{p}_1$or $\mathbf{p}_2$ can be taken a different definition of cell polarity. Using either $\mathbf{p}_1$or $\mathbf{p}_2$ to replace $\mathbf{p}$ does not quantitatively change results in the paper.

**Statistical distribution and analysis**. With the single-cell velocities and polarities calculated above, we obtained microscopic quantities ($\mathbf{u}_{\mathrm{flow}}$; $\mathbf{u}_{\mathrm{cell}}$, $\mathbf{v}_{\mathrm{cell}}$, $\mathbf{p}$) associated with every identifiable individual cells within every image acquisition cycle. Every cell identified within an image acquisition cycle was treated as an independent single-cell event, and the same cell in different acquisition cycles was counted as different events. We note that cells with faster apparent speeds would appear more frequently in the ROI , but they have shorter-duration trajectories in the ROI; as a result, these cells would not be over-sampled in our statistical analysis. Similarly, cells moving along the shorter edge of the ROI have shorter-duration trajectories, but they appear more frequently because of the greater ROI cross-section for these cells (i.e., the longer edge of the ROI). In any case, counting single-cell events in every image acquisition cycle rather than counting cell trajectories across different image acquisition cycles could eliminate the bias due to more frequent appearance but shorter trajectory duration.

We counted all single-cell events in all image acquisition cycles of a time-lapse recording for statistical analysis. For the distribution of an arbitrary angle $\psi$ between two vector directions, we calculated the polar alignment parameter $\langle\cos\psi\rangle$, where $\langle\cdot\rangle$ denotes average over all events in a recording. In particular, $\langle\cos\psi\rangle = 1$ represents perfect parallel alignment and $\langle\cos\psi\rangle = -1$ for perfect anti-parallel alignment, while a uniform distribution of angle $\psi$ leads to $\langle\cos\psi\rangle = 0$.

**Filters in statistical analysis**. For all analysis except when plotting the single-cell speed distributions (Fig. 4a), we filtered out single-cell events in a recording based on three quantities: advection-corrected single-cell speed $v_{\mathrm{cell}} \equiv |\mathbf{v}_{\mathrm{cell}}|$, distance from flagellum center to cell body center $a \equiv |a\mathbf{p}|$, and the angle of the flagellum-to-body direction (polarity) $\mathbf{p}$ with respect to local advection $\mathbf{u}_{\mathrm{flow}}$ (denoted as $\theta$). Here $|\cdot|$ means the magnitude of a vector. Two general filters were applied: 1) Only cells with $v_{\mathrm{cell}}$ in the range of 5-100 μm/s were kept for analysis. Cells with $v_{\mathrm{cell}}$ less than 5 μm/s were considered immotile and cells with $v_{\mathrm{cell}}$ more than 100 μm/s were unphysical and regarded as arising from noise, both of which were not counted in the statistical analysis; in particular, the flagella of immotile cells were passively sheared by the advection flow and their flagellar orientation should not be used for cell polarity calculation. 2) Only cells with $a$ in the range of 1-10 μm were kept for analysis. Cells with $a$ less than 1 μm were either not moving within the observation plane or undergoing tumbles, while cells with $a$ more than 10 μm were unphysical and regarded as arising from noise. Removing the filters did not alter results qualitatively despite minor numerical shifts.

**Robustness of the experimental and data analysis methods.** To examine whether our definition of cell polarity $\mathbf{p}$ as the direction vector from flagellum center to cell body center would obscure any order in flagellar orientation $\mathbf{n}_{\mathrm{fla}}$ (with $\mathbf{n}_{\mathrm{fla}} = -\mathbf{n}_{\mathrm{fla}}$) or cell body orientation $\mathbf{n}_{\mathrm{cell}}$ (with $\mathbf{n}_{\mathrm{cell}} = -\mathbf{n}_{\mathrm{cell}}$), we checked the distributions of angle $\varphi_{\mathrm{fla}}$ (between $\mathbf{n}_{\mathrm{fla}}$ and $\mathbf{u}_{\mathrm{flow}}$) and angle $\varphi_{\mathrm{cell}}$ (between $\mathbf{n}_{\mathrm{cell}}$ and $\mathbf{u}_{\mathrm{flow}}$) (Fig. S10a, Table S2). We found that both angles were uniformly distributed in dense bacterial suspensions at either motile limit (with $r_m = 1.0$) or immotile limit ($r_m = 0.001$) (Fig. S10b-e). Their nematic alignment parameters $Q_{\text{cell}}$ and $Q_{\text{fla}}$ (defined as $Q = \langle 2\cos^2\phi - 1\rangle$, where $\phi$ can be either $\varphi_{\text{cell}}$ or $\varphi_{\text{fla}}$, with $Q = 1$ representing perfect parallel or anti-parallel alignment, $Q = -1$ representing perfectly orthogonal configurations, and $Q = 0$ representing a uniform angle distribution) remained at ~0 regardless of motile cell proportion $r_m$ (Fig. S10f). Therefore, the orientations of flagellar bundles and cell bodies were also randomly distributed with respect to the solvent flows.

We also examined whether the main results presented in main text figures depend on experimental settings (position of imaging plane and cell density) or data processing parameters. First, the results in main text Figs. 2-4 remained unchanged if we chose the imaging plane as ¼ chamber height from the bottom glass surface instead of the mid-plane despite a higher noise. Second, increasing the cell density from ~$8 \times 10^{10}$ cells/ml to ~$1.4 \times 10^{11}$ cells/ml (which is deep within the so-called active turbulence regime (36)) yielded similar results in terms of random cell polarity distribution and upstream navigation. Lastly, the results in main text Figs. 2-4 do not depend on the search window size for PIV analysis or on the specific interpolation scheme for velocity calculation.

**Note on the treatment of cells with out-of-plane motion.** In our imaging setup, successful tracking of cells requires a cell to remain in the focal plane (~1 µm depth) within one acquisition cycle (50 ms), which excludes cells having a large vertical velocity component (>20 µm/s). Meanwhile, as described in section “Filters in statistical analysis” above, we filtered out single-cell events in a recording based on the distance from flagellum center to cell body center $\mathrm{a} \equiv |\mathrm{a}\mathbf{p}|$. Only cells with $\mathrm{a}$ in the range of 1-10 µm were kept for analysis. Cells with $\mathrm{a} < 1$ µm were either not moving within the observation plane or undergoing tumbles. This filter excludes cells whose vertical component of cell polarity is much larger than the in-plane components.

**Hydrodynamic modeling of dense bacterial active fluids**

We perform large-scale simulations of the hydrodynamic interactions in 3D dense suspensions of flagellated bacteria at room temperature, confined by two infinitely large parallel plates with no-slip boundary condition (Fig. 6a). In this simulation, we use $1\ \mu\text{m}$ and $1\ s$ as the unit of length and time respectively, and set the viscosity of water $\mu \approx 10^{-3}\text{N}\cdot\text{s/m}^2$ to be 1 for the unit of force. Accordingly, the two plates are described as $z_1 = 0$ and $z_2 = 40$, respectively, and room temperature of $T = 300K$ can be described as $k_\text{B}T \approx 4$. In each simulation run, $N = 2000$ bacteria are placed in a simulation box of $0 \le x \le L = 80$, $0 \le y \le L$, and $z_1 \le z \le z_2$ with random locations and orientations, where periodic boundary condition is adopted for both $x$ and $y$ directions. The data presented in Fig. 6 are obtained as the average of 10 independent runs, where in each run we simulate the dynamics for 4000 time steps and use the configurations in the last 2000 steps for analysis.

For active suspensions of *E. coli* bacteria, it has been demonstrated in our previous studies (47, 48) that the elongated microswimmer consisting a body and a flagellar bundle can be effectively modelled as a two-sphere composite system rather than a single-component particle: each bacterium is simplified as two spheres with no-slip boundary condition: one sphere with radius $R_\text{b} = 1$ for the bacterial body and the other with radius $R_\mathrm{t} = 0.4$ for the flagellar bundle, connected by a rigid rod that separates the two spheres with a center-to-center distance of $l = 5$. Furthermore, it is demonstrated that the stress distributed on bacterial surface can be effectively modeled as one point-forces and one point-torques acting on the center of each of the two spheres. In this form, the stresslet moment for the bacterial surface-stress can be naturally and explicitly captured by two point-forces on the two spheres separated by distance $l = 5$, and the localized stresslet on the surface of each individual sphere is considered a much smaller quantity and is thus neglected. By use of this two-sphere composite, we are able to faithfully capture near-field hydrodynamic interaction without the unphysical divergence of power consumption associated with the squirmer model that prescribes a configuration-independent slip surface velocity. This two-sphere composite also captures the natural anisotropic geometry of each *E. coli* bacterium.

For simplicity, we assume that all bacteria are smooth swimmers that do not tumble. The Reynolds number of the system is so low ($10^{-5}$) that bacterial flows can be approximately studied through the linear Stokes equation. Then for the dynamics simulated at a time resolution $\Delta t = 10^{-2}\mathrm{s}$, the system is in the overdamped limit described by

$$\boldsymbol{\xi} \cdot (\boldsymbol{U} - k_\mathrm{B} T \nabla \cdot \boldsymbol{\xi}^{-1}) = \boldsymbol{F}^\mathrm{P} + \boldsymbol{F}^\mathrm{B}. \tag{M1}$$

Here the resistance tensor $\boldsymbol{\xi}$ fully describe the hydrodynamic interactions for any given configuration (thus accounting for both the near-field and far-field hydrodynamic interactions) and can be obtained through the classical Stokesian Dynamics Simulation method (49, 50); $k_\mathrm{B} T \nabla \cdot \boldsymbol{\xi}^{-1}$ is a thermodynamic effect that becomes significant when bacteria are close to each other or close to the no-slip plates (48); $\boldsymbol{U} \equiv (\boldsymbol{U}_1, \dots, \boldsymbol{U}_i, \dots, \boldsymbol{U}_N)^T$ with $\boldsymbol{U}_i \equiv \left(\boldsymbol{u}_{i\mathrm{b}}, \boldsymbol{\omega}_{i\mathrm{b}}, \boldsymbol{u}_{i\mathrm{t}} - \boldsymbol{u}_i^0, \boldsymbol{\omega}_{i\mathrm{t}}\right)^T$ being the translational/rotational velocity vector characterizing the motions of the body-sphere and tail-sphere, respectively, for the $i$th bacterium; $\boldsymbol{F}^\mathrm{P} \equiv \left(\boldsymbol{F}_1^\mathrm{P}, \dots, \boldsymbol{F}_i^\mathrm{P}, \dots, \boldsymbol{F}_N^\mathrm{P}\right)^T$ with $\boldsymbol{F}_i^\mathrm{P} \equiv (\boldsymbol{F}_{i\mathrm{b}}, \boldsymbol{L}_{i\mathrm{b}}, \boldsymbol{F}_{i\mathrm{t}}, \boldsymbol{L}_{i\mathrm{t}})^T$ being the intrinsic, nonhydrodynamic forces/torques for the body-sphere and tail-sphere, respectively, for the $i$th bacterium; and $\boldsymbol{F}^\mathrm{B}$ represents the stochastic forces/torques (thermal effect of the solvent) that satisfies the multi-variance normal distribution characterized by $\langle \boldsymbol{F}^\mathrm{B}(\Delta t) \rangle = 0$ and $\langle \boldsymbol{F}^\mathrm{B}(\Delta t) \boldsymbol{F}^\mathrm{B}(\Delta t) \rangle = 2 k_\mathrm{B} T \boldsymbol{\xi} \Delta t$. In this simulation we choose to neglect this random noise part by setting $\boldsymbol{F}^\mathrm{B} = 0$.

Please note that in conventional Stokesian Dynamics for single-component particles, individual stresslet blocks and rate-of-strain coupling are also explicitly resolved. In our two-sphere formulation, surface stresslets and rate-of-strain fields for each individual sphere are truncated, because the intra-sphere stresslets are much smaller than the leading order stresslet captured by the two point-forces on the two spheres separated by distance $l = 5\mu\mathrm{m}$. Thus in our formulation Eq. (M1), we connect force and torque ($\boldsymbol{F}^\mathrm{P}$) to translational and rotational velocities relative to the ambient flow ($\boldsymbol{U}$) through the corresponding resistance tensor $\boldsymbol{\xi} = \begin{bmatrix} \boldsymbol{\xi}_{11} & \cdots & \boldsymbol{\xi}_{1N} \\ \vdots & \ddots & \vdots \\ \boldsymbol{\xi}_{N1} & \cdots & \boldsymbol{\xi}_{NN} \end{bmatrix}$, where each sub-matrix $\boldsymbol{\xi}_{ij} = \begin{bmatrix} \boldsymbol{\xi}_{ij}^{FU} & \boldsymbol{\xi}_{ij}^{F\omega} \\ \boldsymbol{\xi}_{ij}^{LU} & \boldsymbol{\xi}_{ij}^{L\omega} \end{bmatrix}$ contains only the $\boldsymbol{\xi}^{FU}$, $\boldsymbol{\xi}^{F\omega}$, $\boldsymbol{\xi}^{LU}$, $\boldsymbol{\xi}^{L\omega}$ blocks that connects force to translational velocity, force to rotational velocity, torque to translational velocity, torque to rotational velocity, respectively.

The evolution of the system can be simulated based on Eq. (M1), where we get resistance tensor $\boldsymbol{\xi}$ through a two-step procedure following the classical Stokesian Dynamics Simulation method (49, 50). In the first step we obtain the far field mobility tensor through $\boldsymbol{M}_{\infty} = \boldsymbol{M}_0 + \widehat{\boldsymbol{M}}$ where $\boldsymbol{M}_0$ is the analytic far-field hydrodynamic interaction without plane (56), and the plane contribution $\widehat{\boldsymbol{M}}$ can be obtained through the method of images (50). In the second step, $\boldsymbol{\xi}$ is obtained through $\boldsymbol{\xi} = \boldsymbol{M}_{\infty}^{-1} + \boldsymbol{\xi}_{\mathrm{near}}$, with $\boldsymbol{\xi}_{\mathrm{near}} = \boldsymbol{\xi}_{\mathrm{2b}} + \boldsymbol{\xi}_{\mathrm{w}}$ describing the analytic lubrication between any two nearby spheres ($\boldsymbol{\xi}_{\mathrm{2b}}$), and between one sphere and the nearby wall ($\boldsymbol{\xi}_{\mathrm{w}}$), respectively. Then multiplying Eq. (1) by $\boldsymbol{M}_{\infty}$ on both sides gives:

$$(\boldsymbol{I} + \boldsymbol{M}_{\infty} \cdot \boldsymbol{\xi}_{\mathrm{near}}) \cdot (\boldsymbol{U} - k_{\mathrm{B}} T \nabla \cdot \boldsymbol{\xi}^{-1}) = \boldsymbol{M}_{\infty} \cdot \left(\boldsymbol{F}^{\mathrm{P}} + \boldsymbol{F}^{\mathrm{B}}\right), \quad \text{(M2)}$$

which is Eq. (3) in the main text. With $\boldsymbol{M}_{\infty}$ and $\boldsymbol{\xi}_{\mathrm{near}}$ analytically available, the system dynamics can be resolved with the application of two widely used conditions. One is the free-swimming condition, in the form of $\boldsymbol{F}_{i\mathrm{b}} = -\boldsymbol{F}_{i\mathrm{t}} \equiv \boldsymbol{F}_i^{\mathrm{eff}}$, $\boldsymbol{L}_{i\mathrm{b}} = -\boldsymbol{L}_{i\mathrm{t}} - (\boldsymbol{r}_{i\mathrm{t}} - \boldsymbol{r}_{i\mathrm{b}}) \times \boldsymbol{F}_{i\mathrm{t}} \equiv -\boldsymbol{L}_i^{\mathrm{eff}} + \boldsymbol{F}_i^{\mathrm{eff}} \times (\boldsymbol{r}_{i\mathrm{b}} - \boldsymbol{r}_{i\mathrm{t}})$. The other is the rigid body condition, in the form of $\boldsymbol{\omega}_{i\mathrm{b}} = \boldsymbol{\omega}_{i\mathrm{t}} = \boldsymbol{\omega}_i^0$ and $\boldsymbol{u}_{i\mathrm{t}} = \boldsymbol{u}_{i\mathrm{b}} + \boldsymbol{\omega}_i^0 \times (\boldsymbol{r}_{i\mathrm{t}} - \boldsymbol{r}_{i\mathrm{b}})$. Then for any given configuration, we can solve Eq. (M2) to obtain $\left\{\boldsymbol{u}_{i\mathrm{b}}, \boldsymbol{\omega}_i^0, \boldsymbol{F}_i^{\mathrm{eff}}, \boldsymbol{L}_i^{\mathrm{eff}}\right\}$ for any $i$th bacterium, where $\boldsymbol{u}_{i\mathrm{b}}$ and $\boldsymbol{\omega}_i^0$ determine the bacterial motion, $\boldsymbol{F}_i^{\mathrm{eff}}$ and $\boldsymbol{L}_i^{\mathrm{eff}}$ describe the force/torque.

The flow field in the suspension at arbitrary location $\boldsymbol{r}_{\mathrm{f}}$ in the solvent regime, can then be obtained by placing a free infinitesimal passive sphere at $\boldsymbol{r}_{\mathrm{f}}$ and computing its motion $\boldsymbol{u}_{\mathrm{f}}$. For this $N$ swimmer plus 1 sphere system, the far field mobility tensor becomes: $\begin{pmatrix} \boldsymbol{M}_{\infty} & \boldsymbol{M}_{\mathrm{sf}} \\ \boldsymbol{M}_{\mathrm{fs}} & \boldsymbol{M}_{\mathrm{ff}} \end{pmatrix}$, where $\boldsymbol{M}_{\mathrm{sf}}$, $\boldsymbol{M}_{\mathrm{fs}}$, and $\boldsymbol{M}_{\mathrm{ff}}$ are analytically available. Since the passive sphere is infinitesimal, we assume that it is far, comparing to its own size, from all swimmers and walls. Thus, Eq. (M2) can be extended to the following:

$$\left(\boldsymbol{I} + \begin{pmatrix} \boldsymbol{M}_{\infty} \cdot \boldsymbol{\xi}_{\mathrm{near}} & \boldsymbol{0} \\ \boldsymbol{M}_{\mathrm{fs}} \cdot \boldsymbol{\xi}_{\mathrm{near}} & \boldsymbol{0} \end{pmatrix}\right) \cdot \begin{pmatrix} \boldsymbol{U} - k_{\mathrm{B}} T \nabla \cdot \boldsymbol{\xi}^{-1} \\ \boldsymbol{u}_{\mathrm{f}} \end{pmatrix} = \begin{pmatrix} \boldsymbol{M}_{\infty} & \boldsymbol{M}_{\mathrm{sf}} \\ \boldsymbol{M}_{\mathrm{fs}} & \boldsymbol{M}_{\mathrm{ff}} \end{pmatrix} \cdot \begin{pmatrix} \boldsymbol{F}^{\mathrm{P}} + \boldsymbol{F}^{\mathrm{B}} \\ \boldsymbol{0} \end{pmatrix}. \quad \text{(M3)}$$

From Eq. (M3) we can get $\boldsymbol{u}_{\mathrm{f}} = \boldsymbol{M}_{\mathrm{fs}} \cdot \left(\boldsymbol{F}^{\mathrm{P}} + \boldsymbol{F}^{\mathrm{B}} - \boldsymbol{\xi}_{\mathrm{near}} \cdot (\boldsymbol{U} - k_{\mathrm{B}} T \nabla \cdot \boldsymbol{\xi}^{-1})\right)$, where $\boldsymbol{M}_{\mathrm{fs}}$, $\boldsymbol{\xi}_{\mathrm{near}}$, and $\boldsymbol{M}_{\infty}^{-1}$ are mobility/resistance terms that are analytically available, $\boldsymbol{F}^{\mathrm{P}}$ and $\boldsymbol{U}$ are obtained by solving Eq. (M2) in the last paragraph.

For each configuration, the flow field in the suspension is then obtained from the method above for a set of locations in the solvent regime, from which we obtain the flow field $\boldsymbol{u}_{\mathrm{flow}}$ at arbitrary locations through interpolation. Combining the flow field results and the motion of bacteria, we accumulate the distribution of the various angles as studied in experiments.

**Supplementary Materials**, including Supplementary Videos, are available in the online version of the paper.

**Data availability.** The data supporting the findings of this study are included within the paper and its Supplementary Materials.

**Code availability.** The custom codes used in this study are available from CUHK Research Data Repository system: https://researchdata.cuhk.edu.hk/dataverse/microscopicdynamics3d/.

**Acknowledgements**. We thank Cristina Marchetti, Igor Aranson, Luca Giomi, Eric Lauga, and Julia M. Yeomans for helpful discussions. Y.W. acknowledges support by National Natural Science Foundation of China (NSFC No. T2425007, Y.W.; No. 12474191 for X. X.), the Research Grants Council of Hong Kong SAR (RGC Ref. No. 14307822, 14309023, RFS2021-4S04 and CUHK Direct Grants, Y.W.) and by Ministry of Science and Technology Most China (No. 2021YFA0910700, Y.W.), as well as support from New Cornerstone Science Foundation through the Xplorer Prize.

**Author Contributions**: Y. Wang designed the study, performed experiments, analyzed and interpreted the data. Y.D. helped with data acquisition and analysis. P.L. and X.X. developed the model for hydrodynamic interactions, performed numerical simulations, and interpreted the numerical results. Y. Wu conceived the project, designed the study, analyzed and interpreted the data. Y.W. and X.X. supervised the project. Y. Wu wrote the paper with all authors' input.

**Author Information**: Reprints and permissions information is available on the journal website. The authors declare no competing financial interests. Inquiries on the hydrodynamic simulations should be addressed to X.X. (xinliang@hainanu.edu.cn). All other inquiries and requests should be addressed to Y.W. (ylwu@cuhk.edu.hk).

## References

1. M. C. Marchetti *et al.*, Hydrodynamics of soft active matter. *Reviews of Modern Physics* **85**, 1143–1189 (2013).
2. H. H. Wensink *et al.*, Meso-scale turbulence in living fluids. *Proceedings of the National Academy of Sciences* **109**, 14308–14313 (2012).
3. A. Creppy, O. Praud, X. Druart, P. L. Kohnke, F. Plouraboué, Turbulence of swarming sperm. *Physical Review E* **92**, 032722 (2015).
4. A. Opathalage *et al.*, Self-organized dynamics and the transition to turbulence of confined active nematics. *Proceedings of the National Academy of Sciences* **116**, 4788–4797 (2019).
5. C. Blanch-Mercader *et al.*, Turbulent Dynamics of Epithelial Cell Cultures. *Physical Review Letters* **120**, 208101 (2018).
6. A. Doostmohammadi, T. N. Shendruk, K. Thijssen, J. M. Yeomans, Onset of meso-scale turbulence in active nematics. *Nature communications* **8**, 1–7 (2017).
7. R. Alert, J. Casademunt, J.-F. Joanny, Active Turbulence. *Annual Review of Condensed Matter Physics* **13**, 143–170 (2022).
8. H. Wioland, F. G. Woodhouse, J. Dunkel, J. O. Kessler, R. E. Goldstein, Confinement Stabilizes a Bacterial Suspension into a Spiral Vortex. *Physical Review Letters* **110**, 268102 (2013).
9. H. Wioland, E. Lushi, R. E. Goldstein, Directed collective motion of bacteria under channel confinement. *New Journal of Physics* **18**, 075002 (2016).
10. K.-T. Wu *et al.*, Transition from turbulent to coherent flows in confined three-dimensional active fluids. *Science* **355** (2017).
11. G. Duclos *et al.*, Spontaneous shear flow in confined cellular nematics. *Nature Physics* **14**, 728–732 (2018).
12. C. Chen, S. Liu, X. Q. Shi, H. Chate, Y. Wu, Weak synchronization and large-scale collective oscillation in dense bacterial suspensions. *Nature* **542**, 210–214 (2017).
13. H. Xu, Y. Wu, Self-enhanced mobility enables vortex pattern formation in living matter. *Nature* 10.1038/s41586-024-07114-8, Doi: 10.1038/s41586–41024–07114–41588 (2024).
14. D. Saintillan, Rheology of Active Fluids. *Annual Review of Fluid Mechanics* **50**, 563–592 (2018).
15. A. Sokolov, I. S. Aranson, Reduction of Viscosity in Suspension of Swimming Bacteria. *Physical Review Letters* **103**, 148101 (2009).
16. H. M. López, J. Gachelin, C. Douarche, H. Auradou, E. Clément, Turning Bacteria Suspensions into Superfluids. *Physical Review Letters* **115**, 028301 (2015).
17. S. Guo, D. Samanta, Y. Peng, X. Xu, X. Cheng, Symmetric shear banding and swarming vortices in bacterial superfluids. *Proceedings of the National Academy of Sciences* **115**, 7212–7217 (2018).
18. V. A. Martinez *et al.*, A combined rheometry and imaging study of viscosity reduction in bacterial suspensions. *Proceedings of the National Academy of Sciences* **117**, 2326–2331 (2020).
19. R. E. Goldstein, J.-W. van de Meent, A physical perspective on cytoplasmic streaming. *Interface Focus* **5**, 20150030 (2015).
20. V. Hakim, P. Silberzan, Collective cell migration: a physics perspective. *Reports on Progress in Physics* **80**, 076601 (2017).
21. A. Sokolov, M. M. Apodaca, B. A. Grzybowski, I. S. Aranson, Swimming bacteria power microscopic gears. *Proceedings of the National Academy of Sciences* **107**, 969–974 (2010).

22. R. Di Leonardo *et al.*, Bacterial ratchet motors. *Proceedings of the National Academy of Sciences* **107**, 9541–9545 (2010).
23. A. Kaiser *et al.*, Transport Powered by Bacterial Turbulence. *Physical Review Letters* **112**, 158101 (2014).
24. M. O. Din *et al.*, Synchronized cycles of bacterial lysis for in vivo delivery. *Nature* **536**, 81–85 (2016).
25. S. P. Thampi, A. Doostmohammadi, T. N. Shendruk, R. Golestanian, J. M. Yeomans, Active micromachines: Microfluidics powered by mesoscale turbulence. *Science Advances* **2**, e1501854.
26. S. Li *et al.*, Self-regulated non-reciprocal motions in single-material microstructures. *Nature* **605**, 76–83 (2022).
27. W. Wang *et al.*, Cilia metasurfaces for electronically programmable microfluidic manipulation. *Nature* **605**, 681–686 (2022).
28. I. S. Aranson, A. Sokolov, J. O. Kessler, R. E. Goldstein, Model for dynamical coherence in thin films of self-propelled microorganisms. *Phys. Rev. E* **75**, 040901 (2007).
29. J. Dunkel *et al.*, Fluid Dynamics of Bacterial Turbulence. *Physical Review Letters* **110**, 228102 (2013).
30. S. Heidenreich, J. Dunkel, S. H. Klapp, M. Bär, Hydrodynamic length-scale selection in microswimmer suspensions. *Physical Review E* **94**, 020601 (2016).
31. D. Saintillan, M. J. Shelley, "Theory of Active Suspensions" in Complex Fluids in Biological Systems: Experiment, Theory, and Computation, S. E. Spagnolie, Ed. (Springer New York, New York, NY, 2015), 10.1007/978-1-4939-2065-5_9, pp. 319–355.
32. S. Liu, S. Shankar, M. C. Marchetti, Y. Wu, Viscoelastic control of spatiotemporal order in bacterial active matter. *Nature* **590**, 80–84 (2021).
33. I. Aranson, Bacterial Active Matter. *Reports on Progress in Physics* (2022).
34. J. Colen *et al.*, Machine learning active-nematic hydrodynamics. *Proceedings of the National Academy of Sciences* **118**, e2016708118 (2021).
35. R. Supekar *et al.*, Learning hydrodynamic equations for active matter from particle simulations and experiments. *Proceedings of the National Academy of Sciences* **120**, e2206994120 (2023).
36. Y. Peng, Z. Liu, X. Cheng, Imaging the emergence of bacterial turbulence: Phase diagram and transition kinetics. *Science Advances* **7**, eabd1240 (2021).
37. T. Kaya, H. Koser, Direct Upstream Motility in <em>Escherichia coli</em>. *Biophysical Journal* **102**, 1514–1523 (2012).
38. D. Cao *et al.*, Giant enhancement of bacterial upstream swimming in macromolecular flows. *arXiv preprint arXiv:2408.13694* (2024).
39. B. O. T. Maldonado *et al.*, Bacterial rheotaxis is enhanced in non-Newtonian fluids. *arXiv: 2408.13692* (2024).
40. A. Zöttl, H. Stark, Nonlinear dynamics of a microswimmer in Poiseuille flow. *Phys Rev Lett* **108**, 218104 (2012).
41. D. Saintillan, M. J. Shelley, Instabilities, pattern formation, and mixing in active suspensions. *Physics of Fluids* **20**, 123304 (2008).
42. T. Ishikawa *et al.*, Energy Transport in a Concentrated Suspension of Bacteria. *Physical Review Letters* **107**, 028102 (2011).
43. A. E. Patteson, A. Gopinath, M. Goulian, P. E. Arratia, Running and tumbling with E. coli in polymeric solutions. *Scientific Reports* **5**, 15761 (2015).
44. S. Kamdar *et al.*, The colloidal nature of complex fluids enhances bacterial motility. *Nature* **603**, 819–823 (2022).

45. E. Lauga, T. R. Powers, The hydrodynamics of swimming microorganisms *Rep Prog Phys* **72**, 096601 (2009).
46. N. Yoshinaga, T. B. Liverpool, From hydrodynamic lubrication to many-body interactions in dense suspensions of active swimmers. *The European Physical Journal E* **41**, 76 (2018).
47. B. Zhang, P. Leishangthem, Y. Ding, X. Xu, An effective and efficient model of the near-field hydrodynamic interactions for active suspensions of bacteria. *Proceedings of the National Academy of Sciences* **118**, e2100145118 (2021).
48. P. Leishangthem, X. Xu, Thermodynamic Effects Are Essential for Surface Entrapment of Bacteria. *Physical Review Letters* **132**, 238302 (2024).
49. J. F. Brady, G. Bossis, Stokesian Dynamics. *Annual Review of Fluid Mechanics* **20**, 111–157 (1988).
50. J. W. Swan, J. F. Brady, Simulation of hydrodynamically interacting particles near a no-slip boundary. *Physics of Fluids* **19** (2007).
51. Z. Liu, W. Zeng, X. Ma, X. Cheng, Density fluctuations and energy spectra of 3D bacterial suspensions. *Soft Matter* **17**, 10806–10817 (2021).
52. K. Qi, E. Westphal, G. Gompper, R. G. Winkler, Emergence of active turbulence in microswimmer suspensions due to active hydrodynamic stress and volume exclusion. *Communications Physics* **5**, 49 (2022).
53. S. Goh, E. Westphal, R. G. Winkler, G. Gompper, Alignment-induced self-organization of autonomously steering microswimmers: Turbulence, clusters, vortices, and jets. *Physical Review Research* **7**, 013142 (2025).
54. L. Turner, R. Zhang, N. C. Darnton, H. C. Berg, Visualization of flagella during bacterial swarming. *J. Bacteriol.* **192**, 3259–3267 (2010).
55. L. H. Cisneros, J. O. Kessler, S. Ganguly, R. E. Goldstein, Dynamics of swimming bacteria: Transition to directional order at high concentration. *Physical Review E* **83**, 061907 (2011).
56. D. J. Jeffrey, Y. Onishi, Calculation of the resistance and mobility functions for two unequal rigid spheres in low-Reynolds-number flow. *Journal of Fluid Mechanics* **139**, 261–290 (1984).

**Figures**

Figure 1

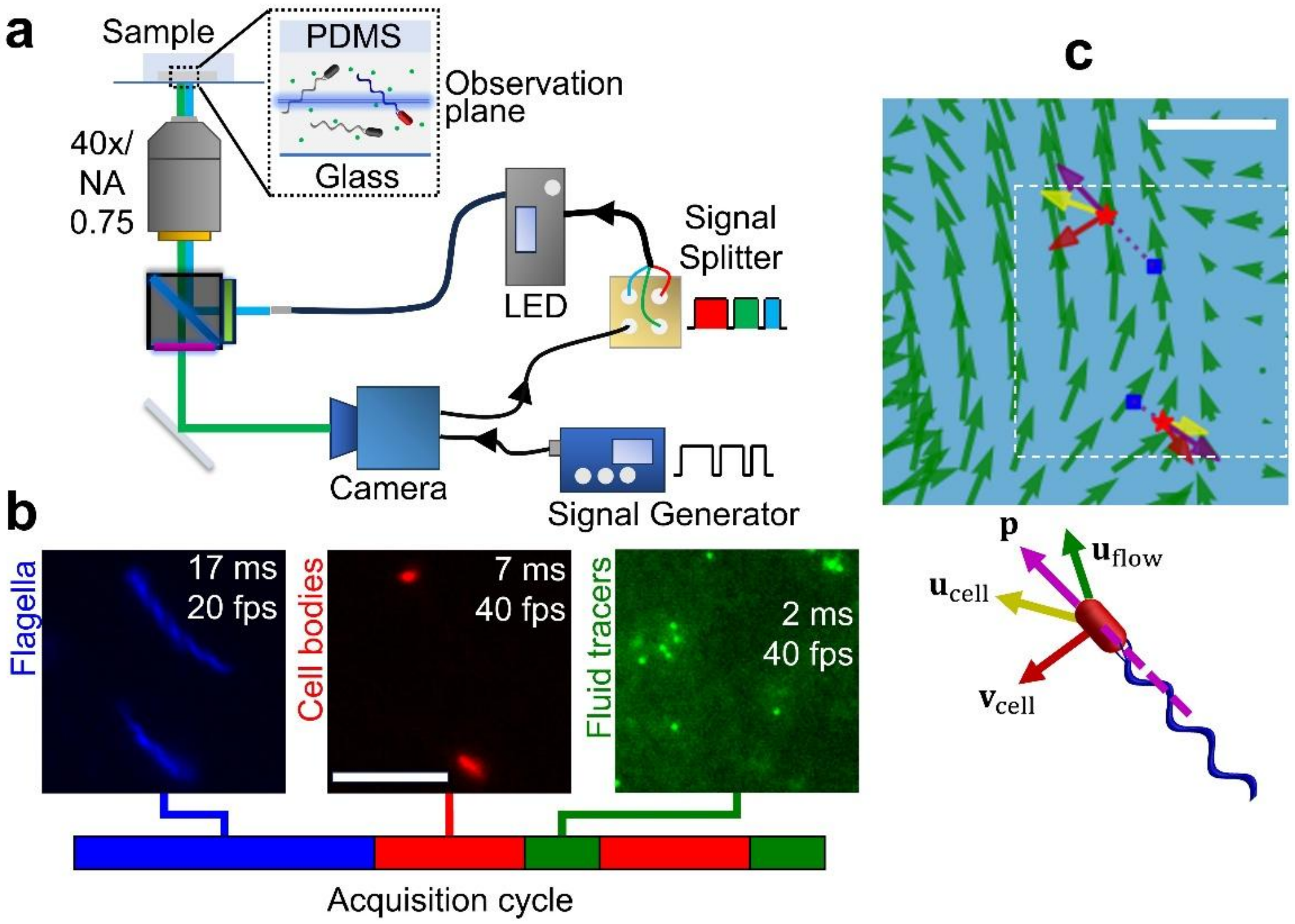

**Fig. 1. Arbitrary-time-sharing multichannel fluorescence imaging for microscopic dynamics measurement in bacterial active fluids.** (a) Overall schematic illustration of the experimental setup. Arrowed lines indicate the control flow of the programmable electronic triggering system. The signal generator sends a triggering signal to the camera to initiate acquisition of each frame and simultaneously, the camera sends an expose-out signal to control the LED excitation wavelength channels via the signal splitter. (b) False-colored raw images of flagella (blue), cell bodies (red), and fluid tracers (green) acquired in one image acquisition cycle. The exposure times and frame rates of each channel used throughout the study are indicated below the schematic but they can be set to arbitrary values. Also see Video 1. (c) Microscopic quantities and flow field measured from the raw images. The dashed box corresponds to the same region as in the raw images of panel b. Blue squares and red stars indicate flagellum centers and cell body centers, respectively. The arrows are color-coded for the following quantities according to the schematic under the plot as follows: purple arrows, cell polarity (i.e., flagellum-to-body direction) $\mathbf{p}$; yellow arrows, apparent single-cell velocity $\mathbf{u}_{\text{cell}}$ (i.e., the cell body velocity); green arrows, local solvent-flow velocity $\mathbf{u}_{\text{flow}}$; red arrows, the advection-corrected single-cell velocity $\mathbf{v}_{\text{cell}} \equiv \mathbf{u}_{\text{cell}} - \mathbf{u}_{\text{flow}}$. Also see Video 2. Scale bars in panels b,c, 10 μm.

Figure 2

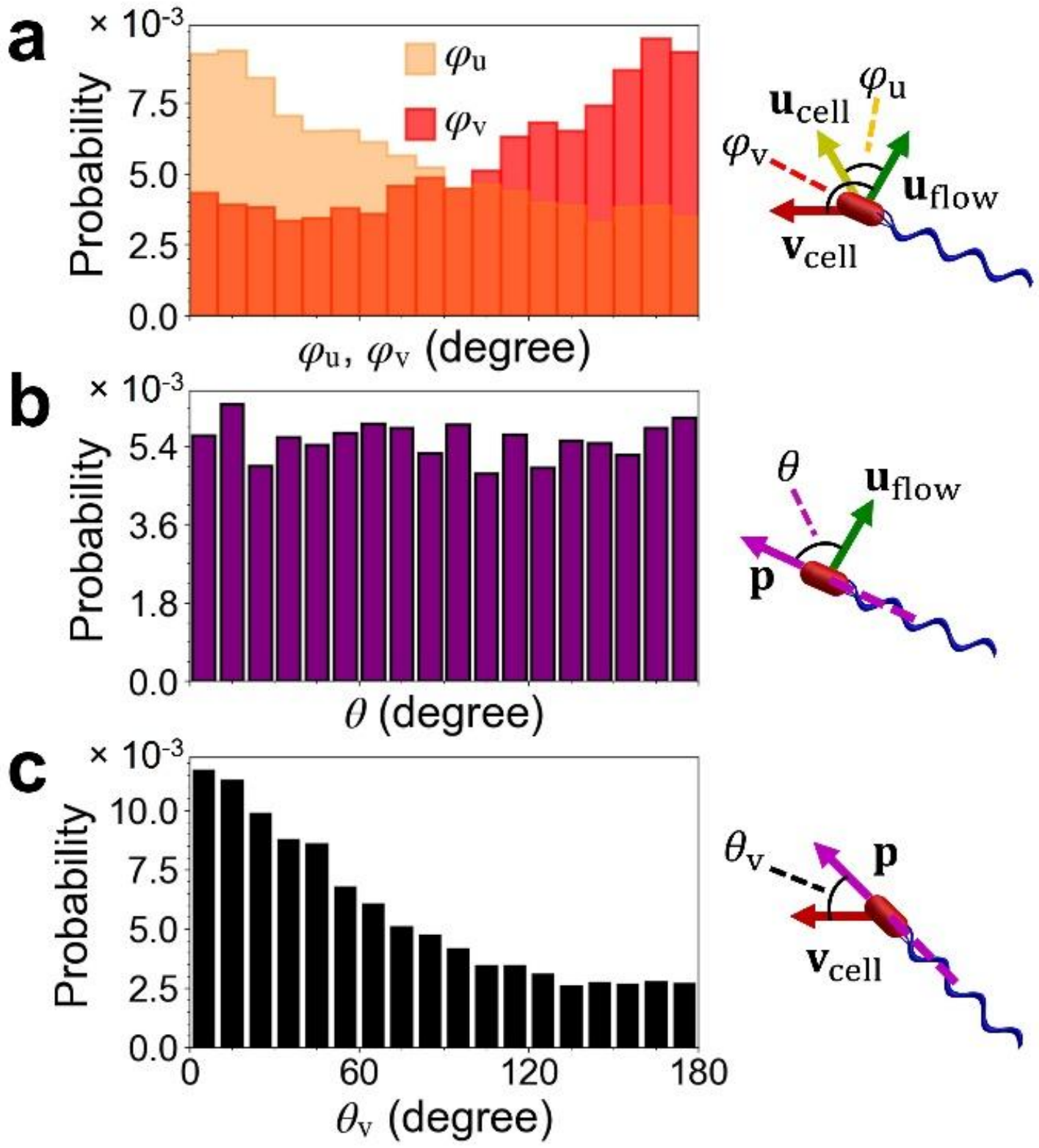

**Fig. 2. Random polarity distribution despite directional bias of cellular transport in bacterial active fluids.** (a) Probability distribution of single-cell velocity directions with respect to local solvent flow. Yellow columns represent the distribution of angle $\varphi_{\mathrm{u}}$ between the apparent single-cell velocity $\mathbf{u}_{\mathrm{cell}}$ and the solvent-flow velocity $\mathbf{u}_{\mathrm{flow}}$, with a downstream bias $\langle\cos\varphi_{\mathrm{u}}\rangle = 0.22$; red columns represent the distribution of angle $\varphi_{\mathrm{v}}$ between the advection-corrected single-cell velocity $\mathbf{v}_{\mathrm{cell}}$ and $\mathbf{u}_{\mathrm{flow}}$, with an upstream bias $\langle\cos\varphi_{\mathrm{v}}\rangle = -0.24$. (b) Probability distribution of angle $\theta$ between cell polarity $\mathbf{p}$ and $\mathbf{u}_{\mathrm{flow}}$, with the $\mathbf{u}_{\mathrm{flow}}$-$\mathbf{p}$ alignment parameter $\langle\cos\theta\rangle \sim 0$. (c) Probability distribution of angle $\theta_{\mathrm{v}}$ between $\mathbf{v}_{\mathrm{cell}}$ and $\mathbf{p}$ with $\langle\cos\theta_{\mathrm{v}}\rangle = 0.36$. Schematic to the right of each panel shows the definition of angles $\varphi_{\mathrm{u}}$, $\varphi_{\mathrm{v}}$, $\theta$, and $\theta_{\mathrm{v}}$. The measurement was performed at cell density 8.4 × $10^{10}$ cells/ml, with the total event count $N = 4440$ (Methods). Three independent experiments reproduced similar distributions.

Figure 3

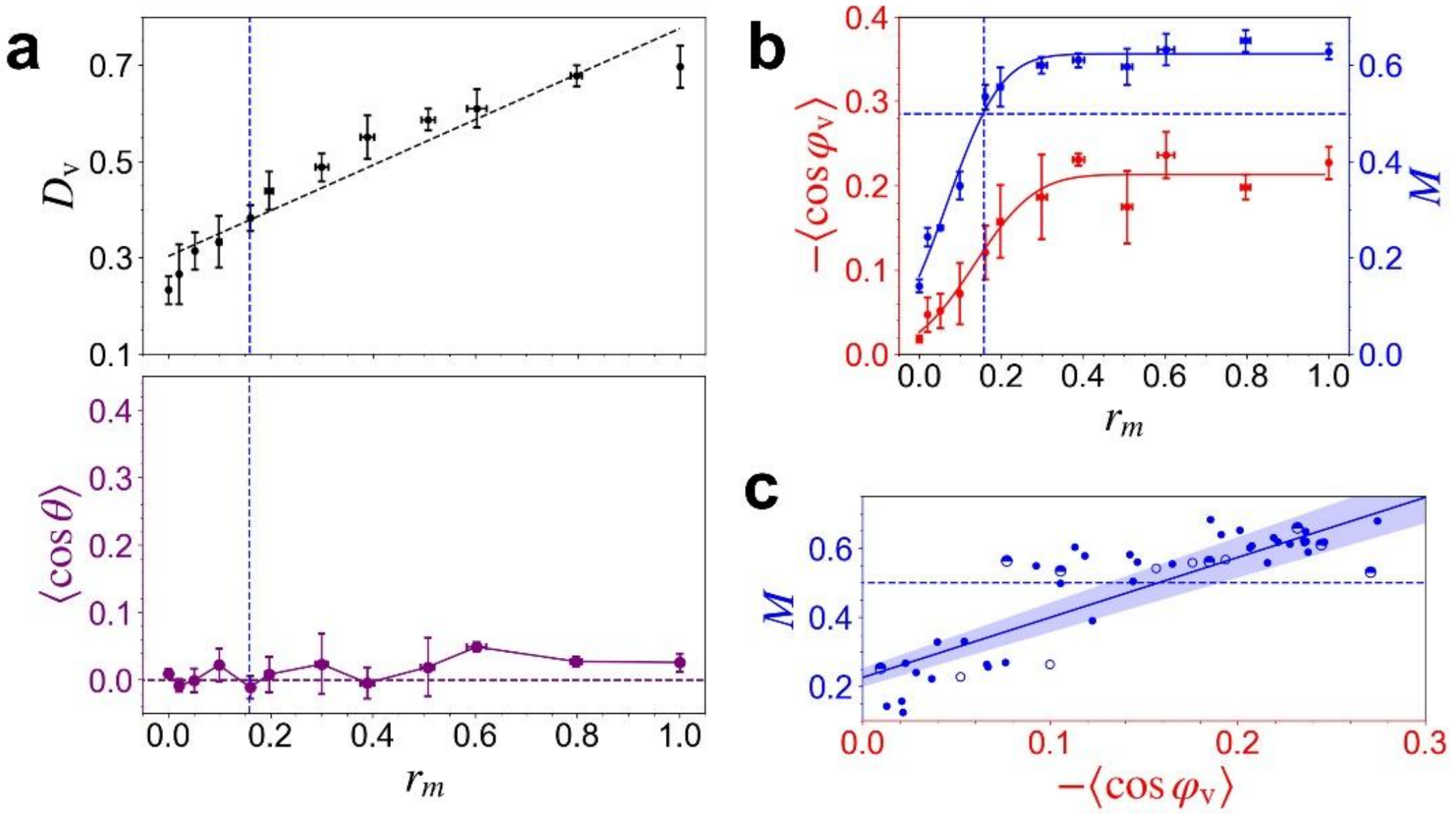

**Fig. 3. Dependence of mesoscale order on hydrodynamic interactions.** (a) Quasi-linear correlation between $\mathbf{v}_{\text{cell}}$-$\mathbf{p}$ misalignment and motile cell proportion $r_m$. Upper panel: $\mathbf{v}_{\text{cell}}$-$\mathbf{p}$ misalignment parameter $D_\text{v} = 1 - \langle \cos\theta_\text{v} \rangle$ plotted against $r_m$; black dashed line is a linear fit, with $R^2 = 0.915$. Lower panel: $\mathbf{u}_{\text{flow}}$-$\mathbf{p}$ alignment parameter $\langle \cos\theta \rangle$ plotted against $r_m$. (b) Upstream navigation index $-\langle \cos\varphi_\text{v} \rangle$ (red) and mesoscale flow order $M$ (blue) plotted against $r_m$. Solid lines are error-function fitting of the experimental data. Total cell density was maintained at ~8 × $10^{10}$ cells/ml. Error bars are standard deviations of 3 independent experiments and the total event count in each experiment $N = 5271 \pm 1136$ (mean±s.d.). (c). Mesoscale flow order $M$ plotted against upstream navigation index $-\langle \cos\varphi_\text{v} \rangle$. Data measured at different total cell densities were presented in the figure: solid circles, ~8 × $10^{10}$ cells/ml (single cell event count for each data point $N$ =5271±1136, mean±s.d.); open circles, ~5 × $10^{10}$ cells/ml (single cell event count for each data point $N$ =5116±1121, mean±s.d.); half-filled circles, >11 × $10^{11}$ cells/ml (single cell event count for each data point $N$ =7024±1809, mean±s.d.). Solid line is a linear fit to all data points ($R^2 = 0.711$, with the shaded area indicating the range within one standard deviation of fitting parameters. In panels a, b, and c, the blue dashed lines indicate the threshold of $r_m$ or $M$ beyond which the dense bacterial suspension displays prominent mesoscale flow order ($M > 0.5$).

Figure 4

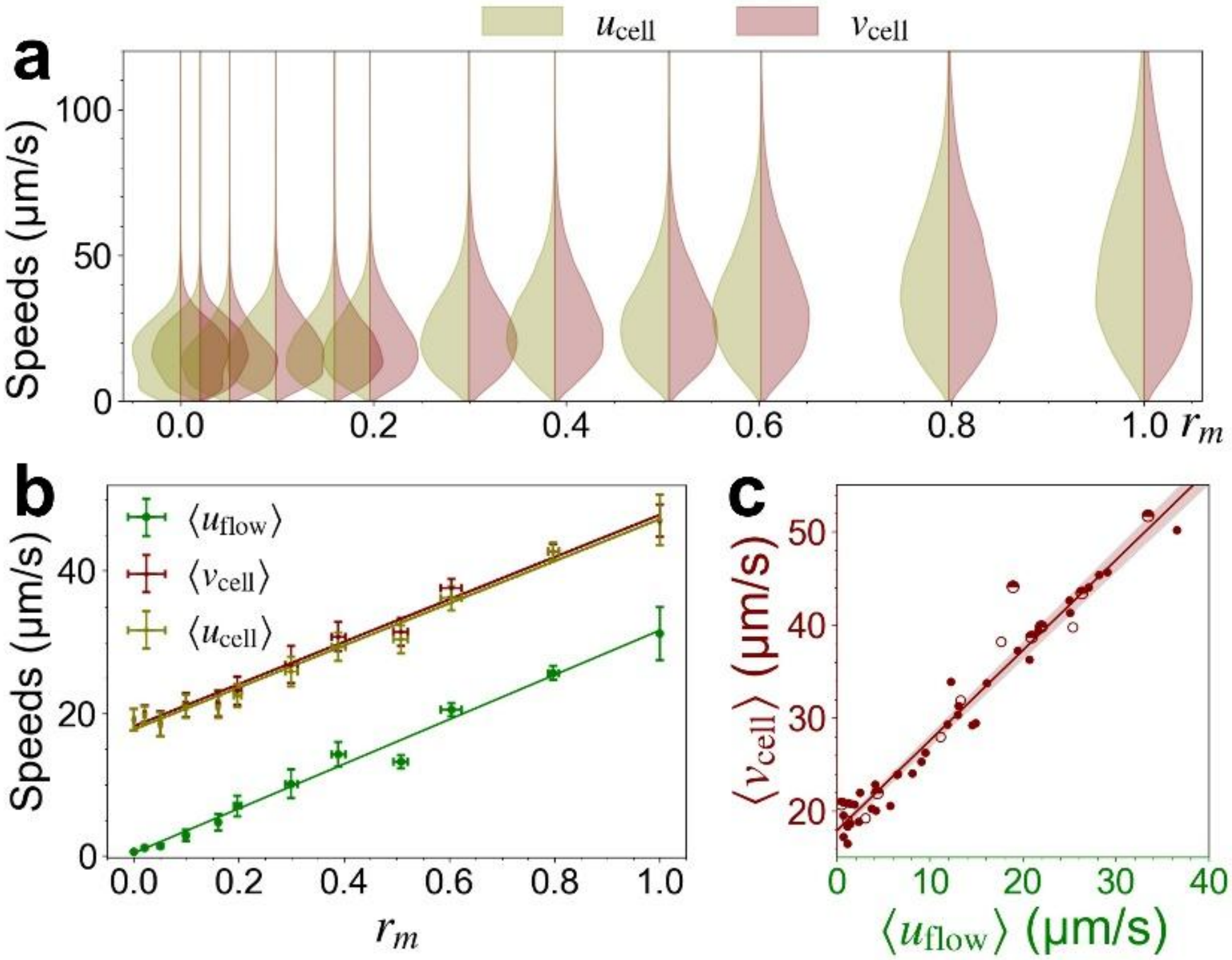

**Fig. 4. Overall enhancement of single-cell speeds controlled by hydrodynamic interactions.** (a) Cellular speed distributions plotted against motile cell proportion $r_m$. Yellow: apparent cellular speed $u_{\text{cell}}$; red: advection-corrected cellular speed $v_{\text{cell}}$. Each violin plot combines data from 3 independent experiments with the same $r_m$ and the total event count in each violin plot $N =$ 15814 ± 2149 (mean±s.d.). (b) Mean cellular speeds plotted against motile cell proportion $r_m$. Yellow: mean apparent single-cell speed $\langle u_{\text{cell}}\rangle$; red: mean advection-corrected single-cell speed $\langle v_{\text{cell}}\rangle$; green: mean solvent-flow speed $\langle u_{\text{flow}}\rangle$. Solid lines are linear fits to the corresponding set of data points ($R^2 > 0.98$). Error bars in panel b represent standard deviations of the mean values of 3 independent experiments (each with a total event count $N = 5271 \pm 1136$, mean±s.d.). Total cell density was maintained at ~8 × $10^{10}$ cells/ml. (c) $\langle v_{\text{cell}}\rangle$ plotted against $\langle u_{\text{flow}}\rangle$ based on data measured at different total cell densities (solid circles, ~8 × $10^{10}$ cells/ml; open circles, ~5 × $10^{10}$ cells/ml; half-filled circles, >11 × $10^{11}$ cells/ml). The solid line is a linear fit to all data points ($R^2 = 0.962$, slope = 0.969 ± 0.001, intercept = 17.9 ± 0.2 μm/s), with the shaded area indicating the range within one standard deviation of fitting parameters.

Figure 5

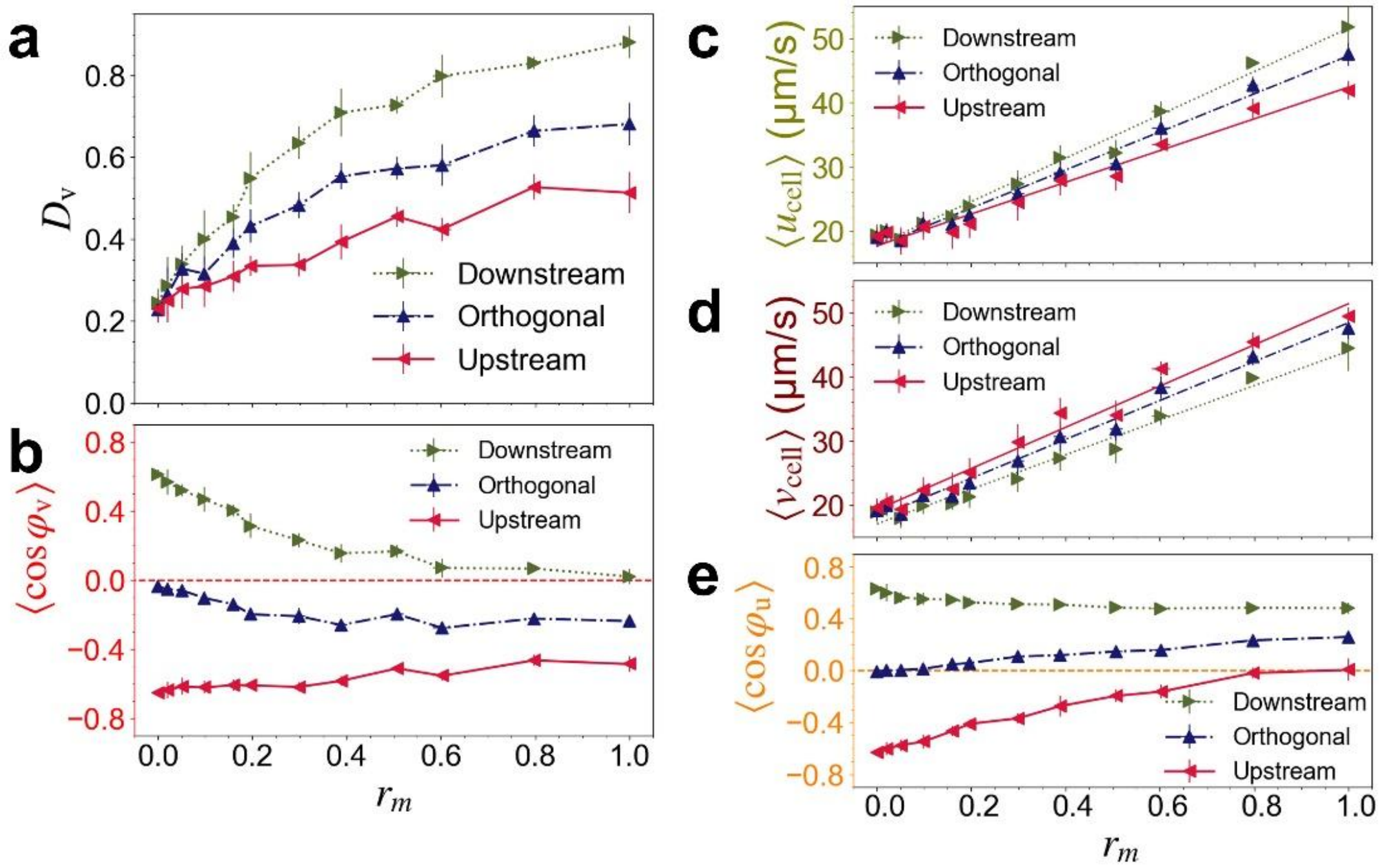

**Fig. 5. Hydrodynamic anisotropy revealed by subpopulation analysis.** The legends shown in panel a apply to all other panels: rightward-pointing triangles in green color for the subpopulation with downstream polarity ($0° \leq \theta < 60°$); upward-pointing triangles in blue color for the subpopulation with orthogonal polarity ($60° \leq \theta < 120°$); leftward-pointing triangles in red color for the subpopulation with upstream polarity ($120° \leq \theta \leq 180°$). The following quantities are plotted against $r_m$ for the three polarity subpopulations: panel a, $\mathbf{v}_{\text{cell}}$-$\mathbf{p}$ misalignment parameter $D_{\text{v}} = 1 - \langle \cos \theta_{\text{v}} \rangle$; panel b, upstream navigation index $\langle \cos \varphi_{\text{v}} \rangle$; panel c, mean apparent single-cell speed $\langle u_{\text{cell}} \rangle$; panel d, mean advection-corrected single-cell speed $\langle v_{\text{cell}} \rangle$; panel e, $\mathbf{u}_{\text{cell}}$-$\mathbf{u}_{\text{flow}}$ alignment parameter $\langle \cos \varphi_{\text{u}} \rangle$, where $\varphi_{\text{u}}$ is the angle between the apparent single-cell velocity $\mathbf{u}_{\text{cell}}$ and the local solvent-flow velocity $\mathbf{u}_{\text{flow}}$ (also see Fig. 2a and Fig. S1e). Lines connecting the data points in panels a,b,e serve as guides for the eye, while lines in panels c,d are linear fits to the corresponding data sets ($R^2 > 0.97$). Error bars in all panels are standard deviations of the mean values of 3 independent experiments (total event count for each experiment $N$ = 1797 ± 379, 1746 ± 387, and 1729 ± 396 for downstream, orthogonal, and upstream polarity subpopulations, respectively). Total cell density in all experiments was maintained at ~8 × $10^{10}$ cells/ml.

Figure 6

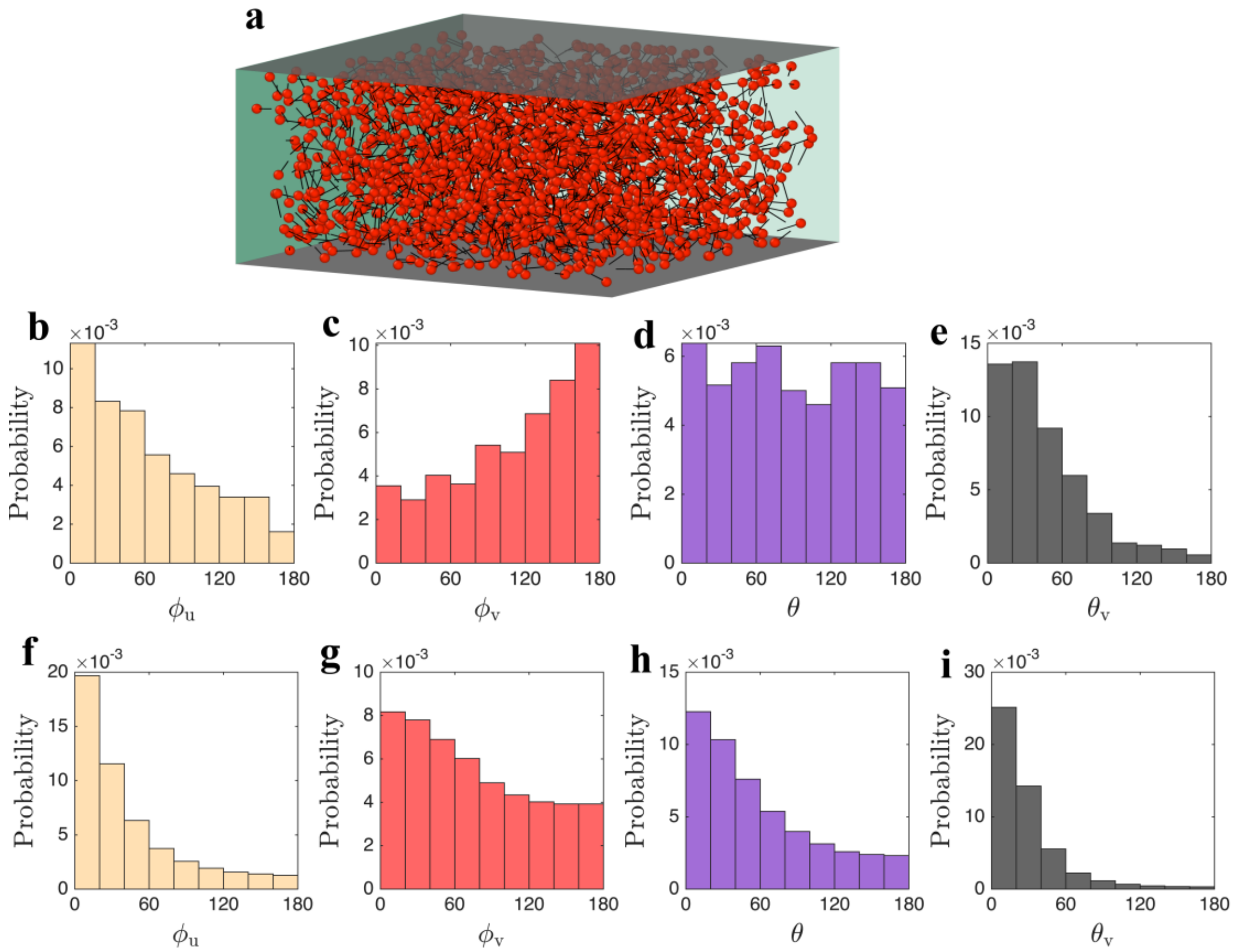

Fig. 6. **Modeling bacterial active fluid with full hydrodynamic interactions**. (a) Setup of the hydrodynamic simulation with $2000$ modelled bacteria in a simulation box of $0 \leq \mathrm{x} \leq \mathrm{L} = 80$, $0 \leq \mathrm{y} \leq \mathrm{L}$, and $0 \leq \mathrm{z} \leq 40$. Also see Methods. (b-e) Probability distributions of $\phi_u$, $\phi_v$, $\theta$, and $\theta_v$ obtained from cells with near neighbors in the simulations, which agrees with the experimental data in Fig. 2a-c. In particular, panel c reproduced the upstream navigation of cells against self-generated solvent flows shown in Fig. 2a, and panel d reproduced the random polarity distribution with respect to local solvent flows in Fig. 2b. (f-i) Probability distributions of $\phi_u$, $\phi_v$, $\theta$, and $\theta_v$ obtained from cells without near neighbors in the simulations.

**SI figures**

Figure S1

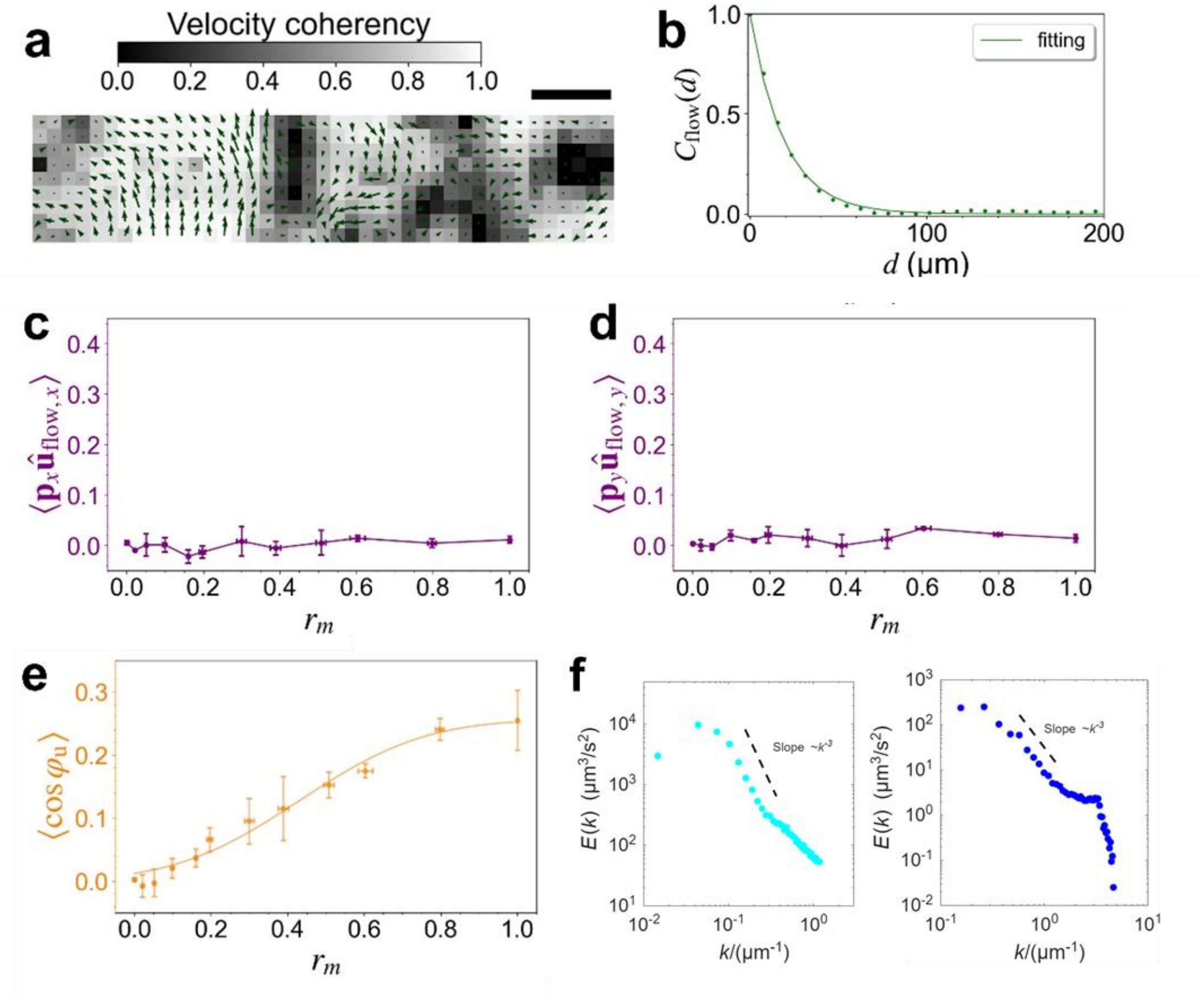

**Fig. S1. Supplementary analysis of single-cell orientational dynamics and flow field in bacterial active fluids.** (a) Spatial distribution of local flow velocity coherency (denoted as $c$; see Methods) in bacterial active fluids. Green arrows represent both the direction and speed of the local solvent flow velocities. Grayscale indicates the value of local velocity coherency associated with each flow velocity vector that measures the local spatial order of the flow velocity field. The mesoscale flow order $M$ ($\in [0,1]$; Methods) is defined as the fraction of velocity vectors with $c > 0.5$ (see Methods); here $M \approx 0.6$. Scale bar, 40 μm. (b) Spatial correlation of solvent flow velocity. The correlation function was calculated as $C_{\mathrm{flow}}(d) = \langle \hat{\mathbf{u}}_{\mathrm{flow}}(\boldsymbol{a}) \cdot \hat{\mathbf{u}}_{\mathrm{flow}}(\boldsymbol{b}) \rangle$, where $\hat{\mathbf{u}}_{\mathrm{flow}}$ is the direction of solvent flow velocity $\mathbf{u}_{\mathrm{flow}}$, $d = |\boldsymbol{a} - \boldsymbol{b}|$ is the distance between the two flow velocity vectors located at $\boldsymbol{a}$ and $\boldsymbol{b}$, and the average $\langle \cdot \rangle$ was performed for all pairs of flow velocity vectors in all image frames in an experiment. Solid line is an exponential fit to the experimental data in the form of $C_{\mathrm{flow}}(d) = \exp(-\frac{d}{d_{\mathrm{flow}}})$. Independent experiments produced similar correlation functions with an average correlation length $d_{\mathrm{flow}}$ of 20.0 ± 1.5 μm (mean ± S.D., N=3 replicates). For panels a,b, the measurement was performed at cell density 7.9 × $10^{10}$ cells/ml (motile cell proportion $r_m = 1$). (c,d) Correlation of the in-plane Cartesian components of cell polarity $\mathbf{p}$ and the direction of self-generated local solvent flow $\hat{\mathbf{u}}_{\mathrm{flow}}$ plotted against motile cell proportion $r_m$. The lag-zero cross-

correlation was calculated here as $\langle \mathrm{p}_x \hat{\mathrm{u}}_{\mathrm{flow},x} \rangle$ and $\langle \mathrm{p}_y \hat{\mathrm{u}}_{\mathrm{flow},y} \rangle$, where the average $\langle \cdot \rangle$ was performed for all cells in an experiment. (e) The $\mathbf{u}_{\mathrm{cell}}$-$\mathbf{u}_{\mathrm{flow}}$ alignment parameter $\langle \cos \varphi_{\mathrm{u}} \rangle$ plotted against motile cell proportion $r_m$. Here $\varphi_{\mathrm{u}}$ is the angle between the apparent single-cell velocity $\mathbf{u}_{\mathrm{cell}}$ and the local solvent-flow velocity $\mathbf{u}_{\mathrm{flow}}$ (see main text Fig. 2a). The solid line is an error-function fitting of the experimental data. This figure shows that, as $r_m$ was increased, cells tended to move downstream in self-generated flows (with $\mathbf{u}_{\mathrm{cell}}$ aligning with $\mathbf{u}_{\mathrm{flow}}$); also see Fig. 5e. For panels c,d,e, the total cell density was maintained at ~$8 \times 10^{10}$ cells/ml, and error bars indicate standard deviation of 3 independent experiments with the total event count in each experiment $N = 15814 \pm 2149$ (mean±s.d.). (f) Kinetic energy spectrum, observed in experiments (light blue; left) and in simulations (dark blue; right), in comparison to decay lines in the form of $k^{-3}$ (dashed lines). The simulation data in small $k$ regime was restricted to $k \sim 2\pi/L \sim 0.1$ due to the limitation of simulation box size. We note that the magnitude of kinetic energy in simulation data is lower than that in the experimental data due to cell density difference.

Figure S2

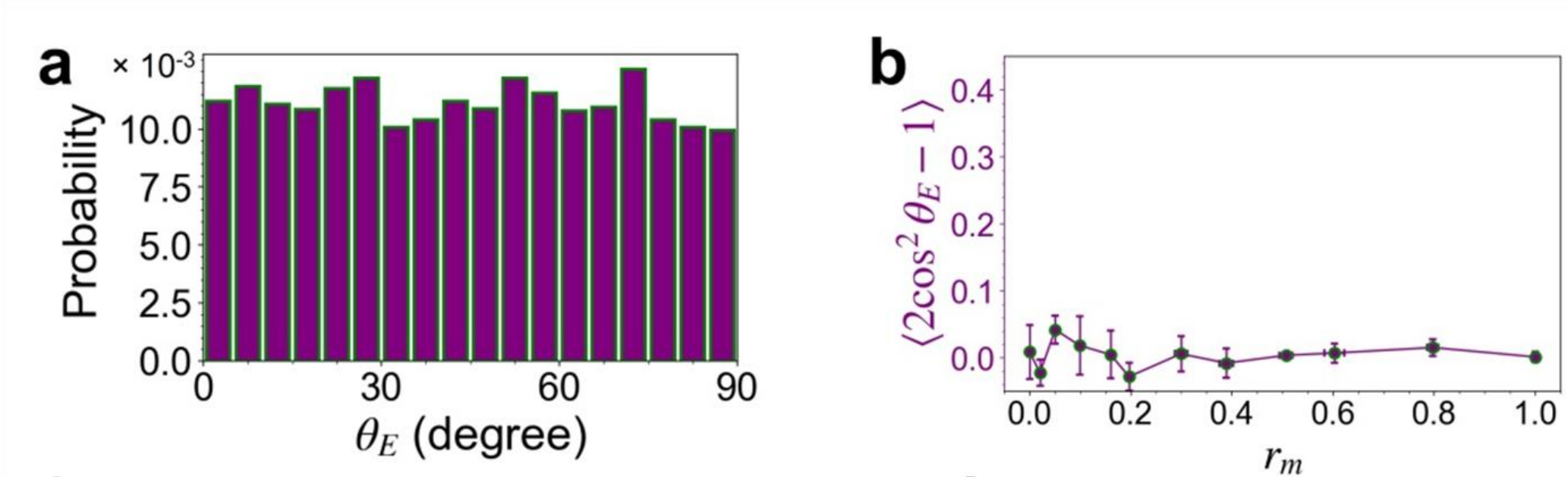

**Fig. S2. Random polarity and cellular transport with respect to the in-plane principal eigenvector of flow strain rate tensor $\boldsymbol{E}$.** (a) Probability distribution of nematic alignment angle $\theta_E$ between cell polarity $\mathbf{p}$ and the in-plane principal eigenvector of $E = (\nabla\mathbf{u}_{\text{flow}}{}^{\text{T}} + \nabla\mathbf{u}_{\text{flow}})/2$, with $\langle\cos 2\theta_E\rangle \sim 0$. The measurement was performed at cell density 8.4 × $10^{10}$ cells/ml, with the total event count $N = 4446$. Three independent experiments reproduced similar distributions. (b) $\langle\cos 2\theta_E\rangle$ plotted against motile cell proportion $r_m$. Total cell density was maintained at ~8 × $10^{10}$ cells/ml. Error bars are standard deviations of 3 independent experiments and the total event count in each experiment $N > 3000$.

Figure S3

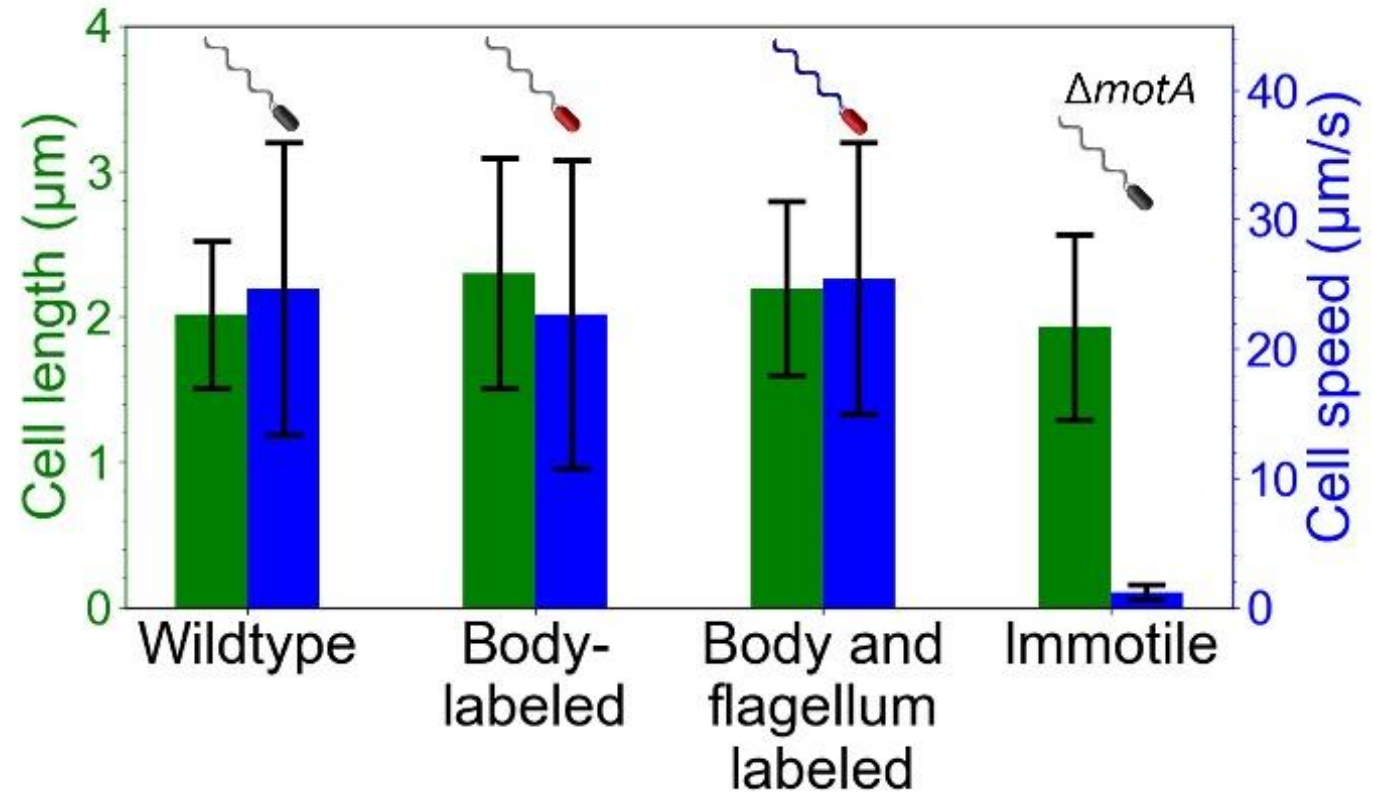

**Fig. S3. Motility and body length of different strains used in the study.** The following *E. coli* strains were used in this study: (1) “Wildtype”, HCB1737 (body length 2.0 ± 0.5 μm, speed 24.6 ± 11.4 μm/s; mean±s.d., total cell number $N_c = 2131$); (2) “Body-labeled”, YW263 constitutively expressing red fluorescent protein in the cytoplasm (body length 2.3 ± 0.8 μm, speed 22.7 ± 11.9 μm/s; mean±s.d., total cell number $N_c = 1173$); (3) “Body and flagellum labeled”, YW263 constitutively expressing red fluorescent protein in the cytoplasm and with it its flagella labeled by Alexa Fluor 647 dye (body length 2.2 ± 0.6 μm, speed 25.4 ± 10.6 μm/s; mean±s.d., total cell number $N_c = 774$); (4) “Immotile”, a Δ*motA* mutant HCB84 (body length 1.9 ± 0.6 μm, speed 1.2 ± 0.6 μm/s; mean±s.d., total cell number $N_c = 356$). Green and blue columns in the plot represent cell body length and cell speed, respectively. Error bars indicate standard deviations from all tracked single cells. Also see Methods for the method of cell tracking and speed calculation. Cells of group 3 “Body and flagellum labeled” were primarily used for the single-cell dynamics measurement in this study. Immotile cells (group 4) differ from other cell groups only in motility. The cell body length and cell swimming speed were measured near a liquid-glass interface in a PDMS-glass chamber filled with a diluted suspension of ~$10^7$ cells/mL. We recorded a two-minute phase-contrast video with a frame rate of 10 fps for cell tracking. Each trajectory that could be tracked continuously was counted as a single cell for cell length and speed statistics; a trajectory was terminated if the cell was stuck to the glass surface. An interval of 0.5 s was used for speed calculation to reduce noise. The cell detection and tracking algorithms were similar to those used in the Methods sections “Recognition of cell body and flagellum” and “Mapping of cell body and flagellum”, respectively. The microscopy setup used here were identical to that in Methods section “Arbitrary-time-sharing multichannel fluorescence microscopy”.

Figure S4

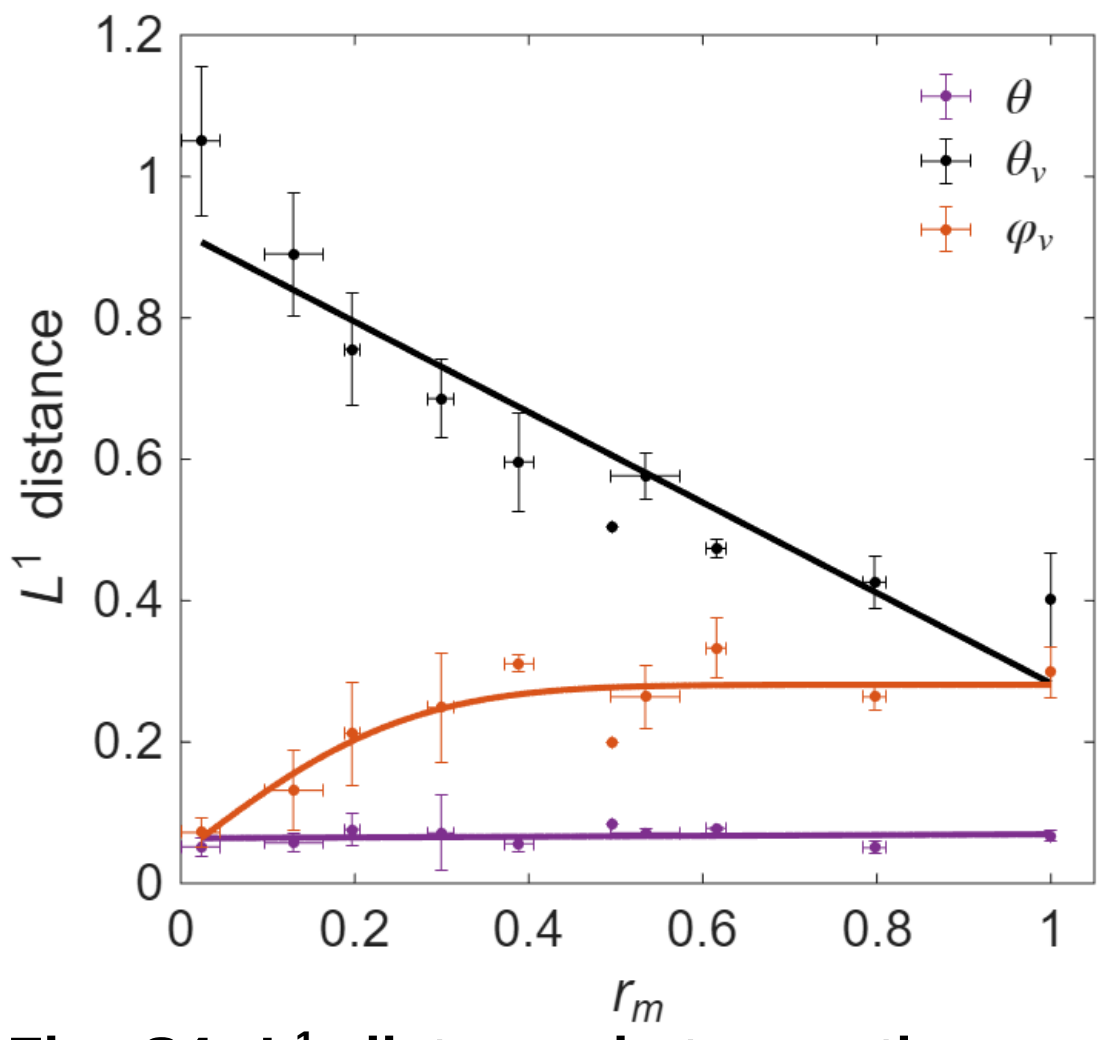

**Fig. S4. L[1] distance between the measured probability distributions of angular quantities and a uniform distribution plotted against $r_m$.** L[1] distance between the measured probability distributions (*P*) and a uniform distribution (*Q*) was calculated as $\sum_i |P_i - Q_i|$, where $P_i$ and $Q_i$ denote the probabilities associated with the $i$-th bin of the two discrete probability distributions *P* and *Q*, respectively. Purple, orange, and black data correspond to the L[1] distance calculated for the probability distributions of $\theta$, $\varphi_v$ and $\theta_v$, respectively; they are consistent with results in Fig. 3a (lower panel), Fig. 3b (red data points), and Fig. 3a (upper panel) of main text, respectively. Total cell density was maintained at ~8 × $10^{10}$ cells/ml. Error bars are standard deviations of 3 independent experiments and the total event count in each experiment >3000.

Figure S5

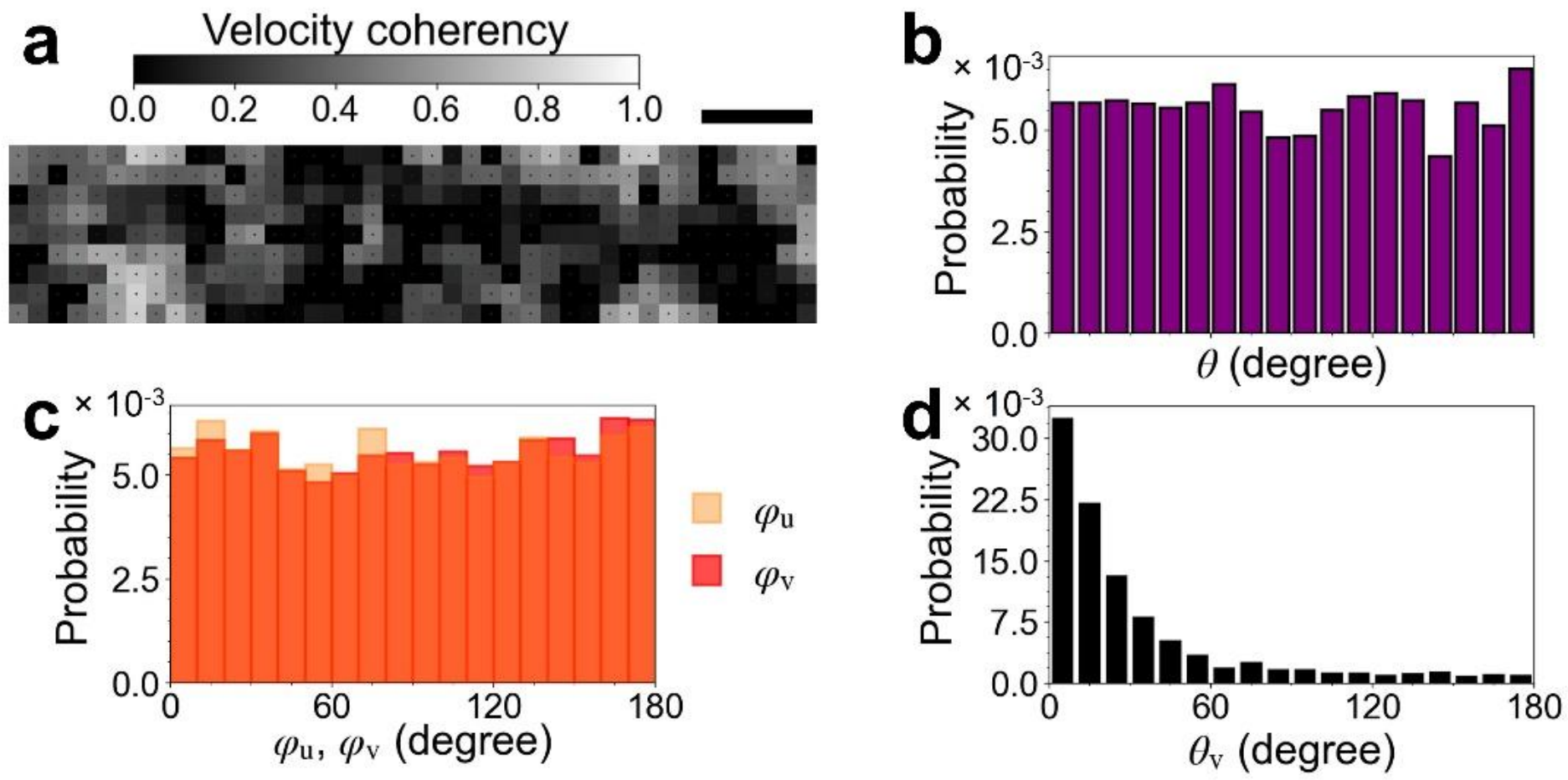

**Fig. S5. Single-cell orientation dynamics in bacterial active fluids in the immotile limit ($r_m \approx 0.001$).** Here the proportion of motile cells (YW263) in the dense bacterial suspension was only ~0.1%. (a) Spatial distribution of local flow velocity coherency in the immotile limit. The figure is plotted in the manner as in Fig. S1b (also see Methods). The mesoscale flow order here is $M \approx 0.1$. Scale bar, 40 μm. (b) Probability distribution of angle $\theta$ between cell polarity $\mathbf{p}$ and local solvent-flow velocity $\mathbf{u}_{\text{flow}}$, with the $\mathbf{u}_{\text{flow}}$-$\mathbf{p}$ alignment parameter $\langle\cos\theta\rangle\sim 0$. (c) Probability distribution of single-cell velocity directions with respect to local solvent flow velocity in the immotile limit. Yellow and red columns represent the distribution of angle $\varphi_{\text{u}}$ and $\varphi_{\text{v}}$, respectively (see main text Fig. 2a for their definitions); from the data we obtained $\langle\cos\varphi_{\text{u}}\rangle\sim 0$ and $\langle\cos\varphi_{\text{v}}\rangle\sim 0$. (d) Probability distribution of angle $\theta_{\text{v}}$ between the advection-corrected single-cell velocity $\mathbf{v}_{\text{cell}}$ and cell polarity $\mathbf{p}$ in the immotile limit. The data showed a sharp peak at $\theta_{\text{v}} = 0$ (i.e., perfect alignment between $\mathbf{v}_{\text{cell}}$ and $\mathbf{p}$), with $\langle\cos\theta_{\text{v}}\rangle = 0.81$; by contrast, $\langle\cos\theta_{\text{v}}\rangle = 0.36$ in the motile limit $r_m \approx 1.0$ (see main text Fig. 2b). The measurement was performed at cell density 8.1 × $10^{10}$ cells/ml (imaging the mid-plane of the fluid chamber), with the total event count $N = 3360$ in panels b-d. Three independent experiments reproduced similar distributions.

Figure S6

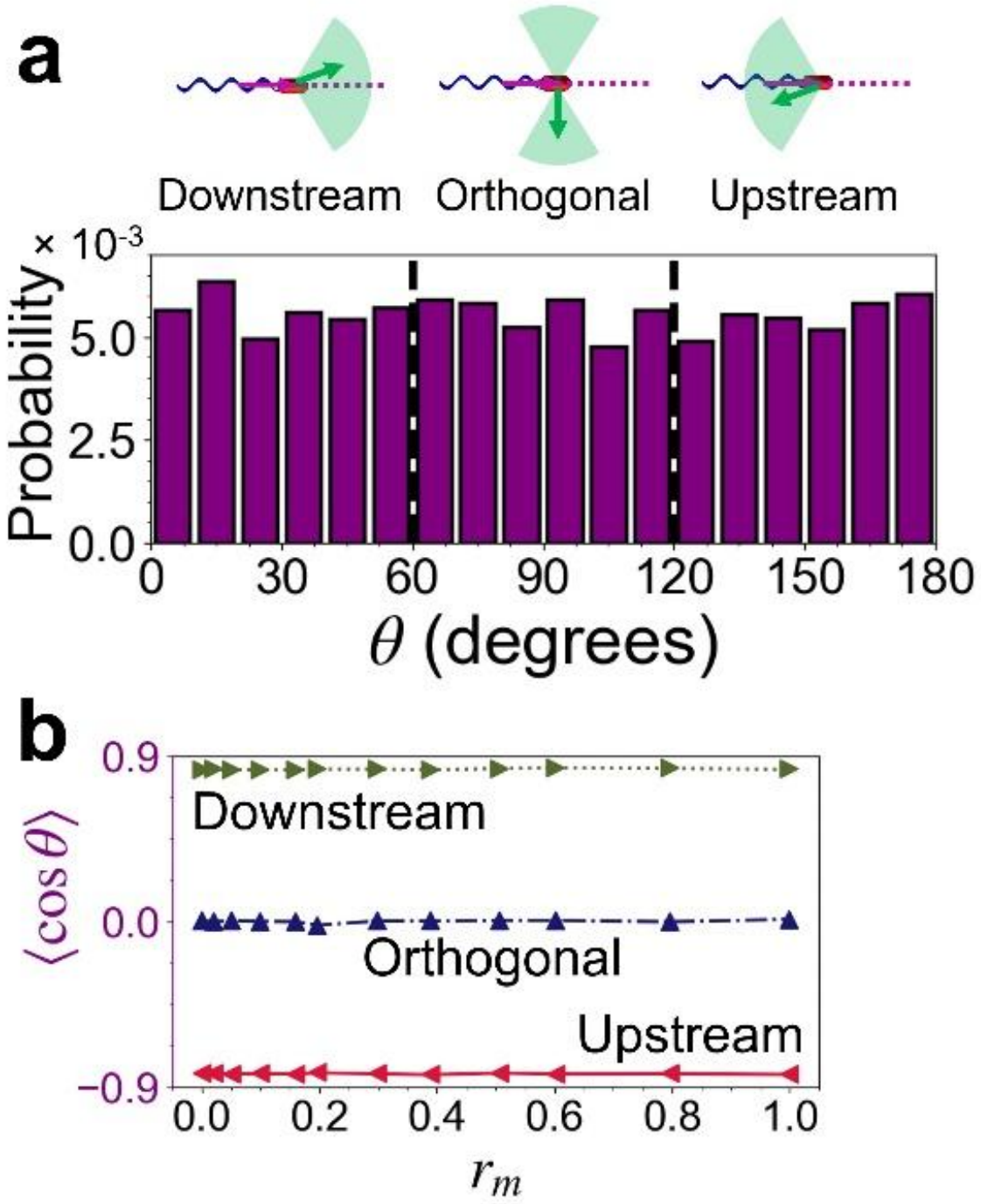

**Fig. S6. Polarity subpopulation analysis.** (a) Schematic illustration of the polarity subpopulation division based on the angle $\theta$ between cell polarity $\mathbf{p}$ and local solvent-flow velocity $\mathbf{u}_{\text{flow}}$. Cells with $\theta = [0°, 60°)$, $\theta = [60°, 120°)$ and $\theta = [120°, 180°]$ belong to the subpopulation with downstream polarity, orthogonal polarity, and upstream polarity, respectively. The probability distribution of angle $\theta$ shown here is identical to that in main text Fig. 2b. (b) $\mathbf{u}_{\text{flow}}$-$\mathbf{p}$ alignment parameter $\langle \cos\theta \rangle$ plotted against motile cell proportion $r_m$ for the three subpopulations. The three polarity subpopulations have $\langle \cos\theta \rangle$ of ~0.9, ~0, and ~ -0.9, respectively. Lines connecting the data points serve as guides for the eye. Error bars (not discernible here due to the small size) are standard deviations of the mean values of 3 independent experiments (total event count in each polarity subpopulation for each experiment $N > 1000$). Total cell density in all experiments was maintained at ~8 × $10^{10}$ cells/ml.

Figure S7

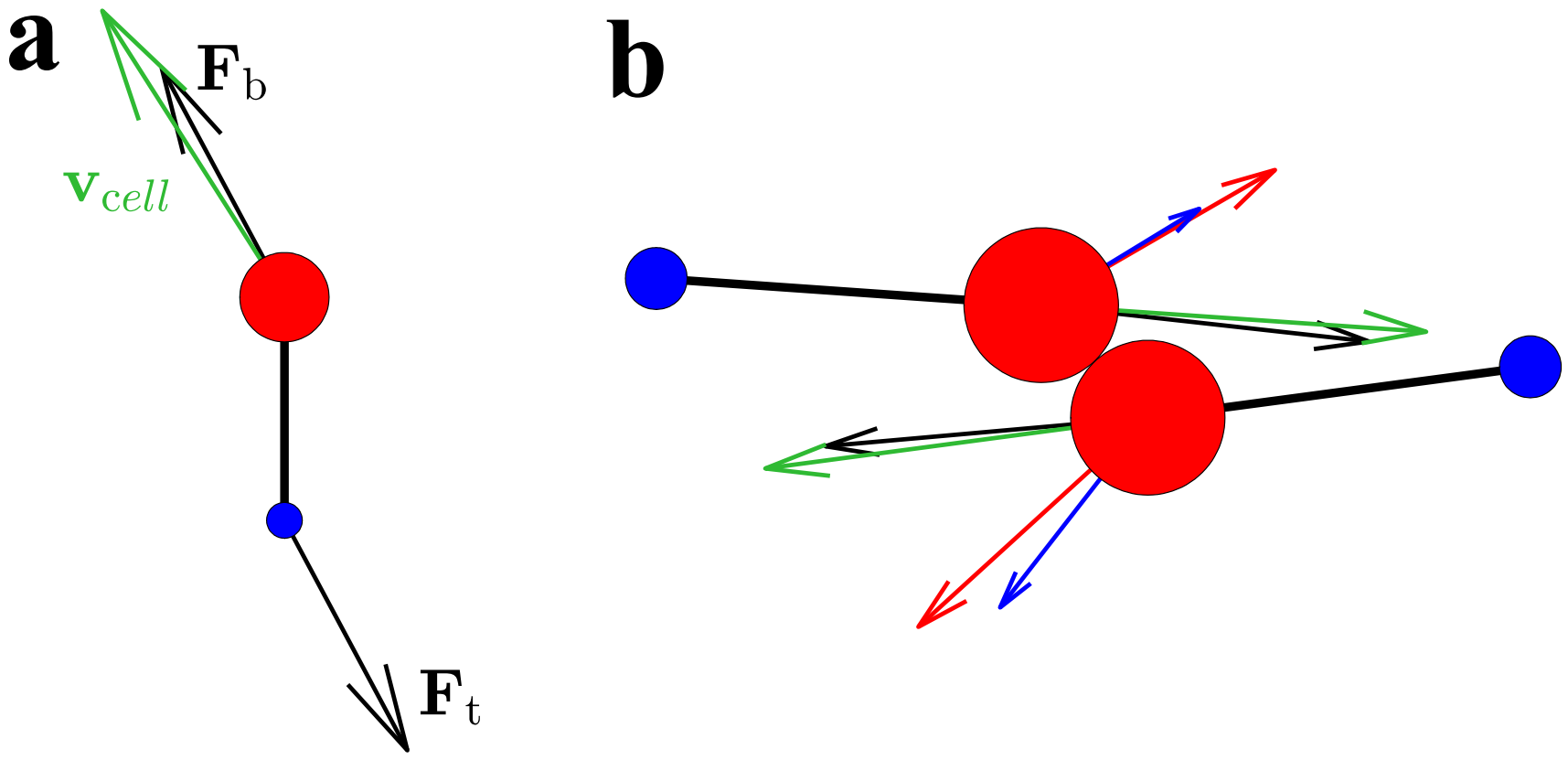

**Fig. S7. Examples showing the misalignment between the bacterial active forcing direction and cell polarity.** (a) A solitary bacterium swimming in a non-uniform flow, where flow on the cell body (red sphere) is towards the right and flow on cell tail (blue sphere) is towards left. The computed intrinsic active forcing ($\boldsymbol{F}_{\mathrm{b}}$ and $\boldsymbol{F}_{\mathrm{t}}$, indicated by black arrows) exerted by the cell to the surrounding fluid and the self-advection velocity $\boldsymbol{v}_{\mathrm{cell}}$ (green arrow) do not align with cell polarity $\boldsymbol{p}$. (b) Two nearby bacteria (red sphere for bacterial body and blue sphere for flagellum) in otherwise quiescent fluid. The computed intrinsic active forcing $\boldsymbol{F}^{\mathrm{P}}$ (black arrows) and bacterial velocities $\boldsymbol{u}_{\mathrm{cell}}$ (red arrows), as well as the total active forcing that generates the flow field $\boldsymbol{F}^{\mathrm{P}} - \boldsymbol{\xi}_{\mathrm{near}} \cdot (\boldsymbol{U} - k_{\mathrm{B}}T\nabla \cdot \boldsymbol{\xi}^{-1})$ (blue arrows), do not necessarily align with cell polarity $\boldsymbol{p}$ (green arrows).

Figure S8

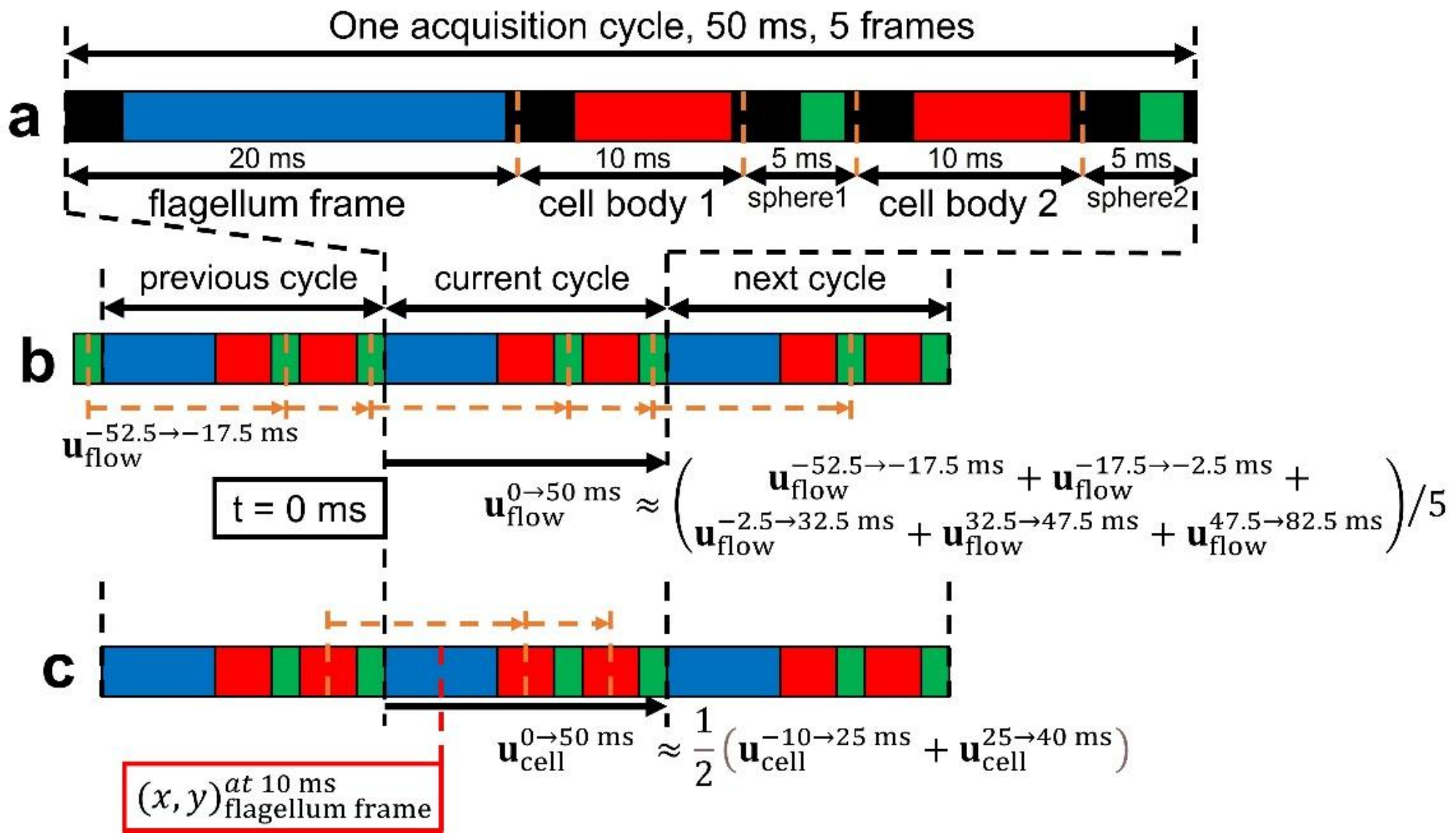

**Fig. S8. Image acquisition cycle in arbitrary-time-sharing multichannel fluorescence microscopy.** (a) Illustration of an image acquisition cycle with a total 50 ms duration. Blue, red, and green bars represent the camera exposure durations of image acquisition for flagellum, cell body, and fluid tracer (microsphere), respectively; each black bar of 3 ms duration in between the colored bars represents the duration of ~2.5 ms expose-out delay (between the trigger-in signal received by the camera from a signal generator and the expose-out signal modulated and sent by the camera to control light source, to ensure that every line of image is exposed for the same duration; it depends on ROI and the mode of expose-out ) and 0.5 ms interval (set in the signal generator). Dashed lines (orange color) represent the time of rising edge of each trigger-in signal. The effective frame rate of imaging is 20 fps (flagella), 40 fps (cell body), and 40 fps (microspheres). Also see Methods for more details. (b) Schematic of calculating local solvent-flow velocity $\mathbf{u}_{\text{flow}}$. To calculate $\mathbf{u}_{\text{flow}}$, microsphere imaging data from 4 consecutive image acquisition cycles is required; panel a is the zoomed-in view of each cycle. PIV analysis was performed for every adjacent pair of the 6 microsphere images in a row: denoting the starting point of the current image acquisition cycle as t = 0 ms, these images were acquired at t = -52.5, -17.5, -2.5, +32.5, +47.5, +82.5 ms, respectively, as indicated by the vertical dashed lines. Through the PIV analysis we obtained velocity field data of $\mathbf{u}_{\text{flow}}$ for 5 time windows, and the average of these 5 velocity fields was taken as the noise-reduced solvent-flow velocity $\mathbf{u}_{\text{flow}}$ for the current image acquisition cycle (from t = 0 ms to 50 ms), which would be used in all subsequent analysis. (c) Schematic of calculating apparent cell velocity $\mathbf{u}_{\text{cell}}$. To calculate $\mathbf{u}_{\text{cell}}$ for the current image acquisition cycle (from t = 0 ms to 50 ms), cell-body positions tracked from 2 consecutive image acquisition cycles (current cycle and previous cycle) are required, and the $\mathbf{u}_{\text{cell}}$ was calculated as the average of cell body velocities tracked in the two time windows [-10 ms, +25 ms] and [+25 ms, 40 ms]. The cell body center position during flagellum imaging, denoted as $(x, y)$, was computed once per cycle by interpolation at t = +10 ms (see Fig. S9n,o and its caption for details); this interpolated cell body center position is for polarity measurement, hence the temporal resolution of polarity data is 20 fps.

Figure S9

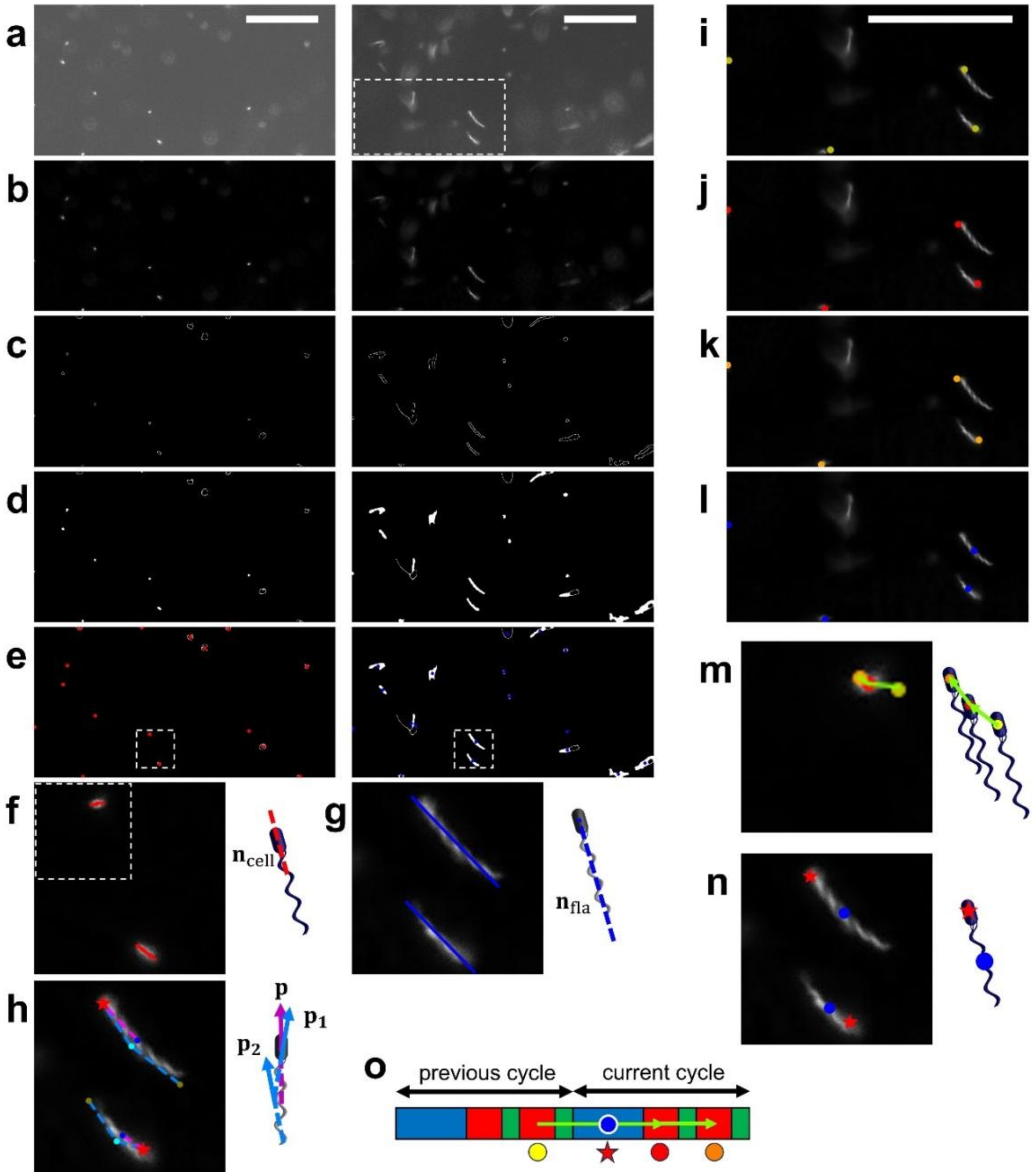

**Fig. S9. Image recognition and mapping of cell bodies and flagella.** (a-d) Each of these panels consists of two image frames, one for the cell body (left) and the other for the flagellum (right). Panels a-d correspond to image processing in the following sequential order: original images (panel a), background normalization and denoising (panel b), edge detection (panel c), and morphological processing of edges by first dilation and then erosion (panel d). (e) Cell body center positions (left) and flagellum center positions (right) were located by labeling the connected regions in panel d (Methods). (f) Recognition of cell-body long axis $b\mathbf{n}_{\text{cell}}$ (red segments). The schematic to the right shows the definition of $\mathbf{n}_{\text{cell}}$. (g) Recognition of flagellum long axis $f_1\mathbf{n}_{\text{fla}}$ (blue segments). The schematic to the right shows the definition of $\mathbf{n}_{\text{fla}}$. (h) Recognition of flagellum center-to-body center vectors (purple dashed lines, $a\mathbf{p}$), flagellum vertex-to-body center vectors (light blue dashed lines, $a_1\mathbf{p}_1$), and flagellum tail-to-vertex vectors (also light blue dashed lines,$a_2\mathbf{p}_2$). Blue dots are flagellum center positions, cyan cots are flagellum vertex positions, yellow dots are flagellum tail end positions, and red stars

are interpolated cell body positions. The schematic to the right shows the definition of $\mathbf{p}$, $\mathbf{p}_1$, and $\mathbf{p}_2$. Features analyzed in panels e-h are based on processed images in panel d. (i-k) Mapping of cell-body center positions between the current image acquisition cycle and the previous acquisition cycle for cell tracking. The cell-body center positions in the first cell-body image frame of the current acquisition cycle (at the mid-time point t = +25 ms; denoting the starting point of the current cycle as t = 0 ms) is shown in panel j (red dots; located by procedures in panels a-e); these positions are mapped or linked to the cell-body center positions in the second cell-body image frame of the previous acquisition cycle (yellow dots in panel i; at t = -10 ms) and in the second cell-body image frame of the current acquisition cycle (orange dots in panel k; at t = +40 ms). The cell-body center positions in panels i-k are overlaid to the flagellum image frame of the current acquisition cycle, which corresponds to the region enclosed by the dashed box in panel a and serves as a reference. (l) Mapping between a cell's flagellum center position and its body center position in the current image acquisition cycle. The cell-body center positions at t = +25 ms (red dots in panel j) are mapped to the flagellum center positions in the current acquisition cycle acquired at t = 10 ms (blue dots). Note that some recognized cell-body or flagellum center positions failed to be mapped or linked together, and these positions are discarded (see Methods). (m) Cell-body displacements (light green lines) based on mapping results in panels i-k. See the schematic to the right. (n) Mapping between a cell's flagellum center position and its interpolated body center position (red star) at the time of flagellum image acquisition (t = 10 ms). The interpolation of cell-body center positions is based on all the position information in panels i-k. See the schematic to the right. (o) Schematic summarizing the feature recognition, mapping, and displacement calculation described in panels i-n. The images in panels f-h and n correspond to the region enclosed by the dashed boxes in panel e. The image in panel m corresponds to the region enclosed by the dashed box in panel f. Scale bars are all 40 μm: panels a-e share the same scale bars in left and right images, respectively; panels i-l share the same scale bar.

Figure S10

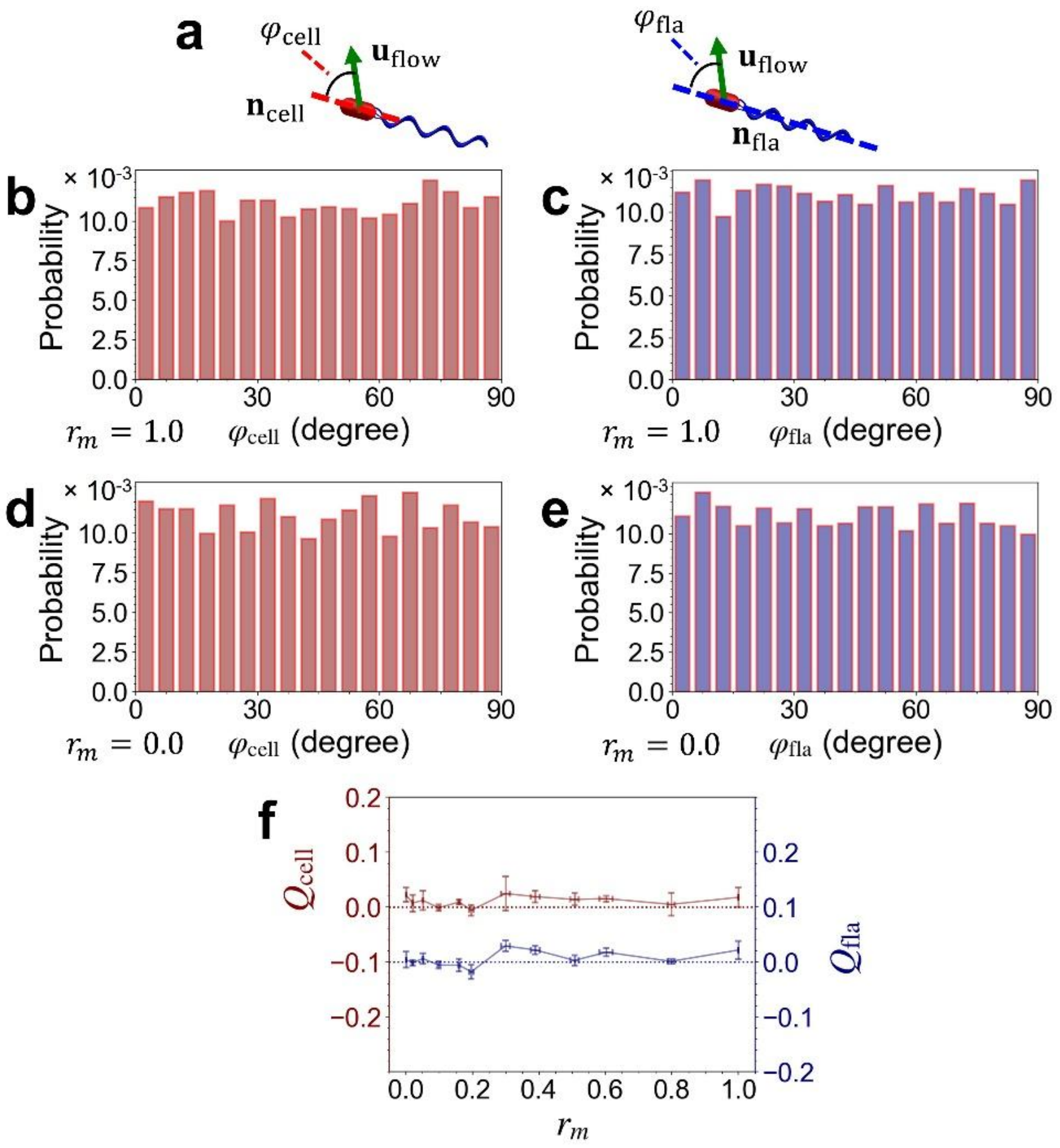

**Fig. S10. Random distribution of cell-body and flagellar bundle orientations.** (a) Schematics showing the definition of the angles $\varphi_{\text{cell}}$ (between cell body long axis $\mathbf{n}_{\text{cell}}$ and local solvent-flow velocity $\mathbf{u}_{\text{flow}}$) and $\varphi_{\text{fla}}$ (between flagellum long axis $\mathbf{n}_{\text{fla}}$ and local solvent-flow velocity $\mathbf{u}_{\text{flow}}$). Note the symmetry $\mathbf{n}_{\text{cell}} = -\mathbf{n}_{\text{cell}}$ and $\mathbf{n}_{\text{fla}} = -\mathbf{n}_{\text{fla}}$. Only cells with an aspect ratio >1.5 were considered in the analysis of $\mathbf{n}_{\text{cell}}$ in order to reduce errors (Methods). (b,d) Probability distributions of angle $\varphi_{\text{cell}}$ at $r_m = 1.0$ (panel b) and at $r_m = 0.001$ (panel d). Both distributions showed a nematic alignment parameter $Q_{\text{cell}}{\sim}0$. (c,e) Probability distributions of angle $\varphi_{\text{fla}}$ at motile cell proportion $r_m = 1.0$ (panel c) and at $r_m = 0.001$ (panel e). Both distributions showed a nematic alignment parameter $Q_{\text{fla}}{\sim}0$. For panels b,c the measurement was performed at cell density 7.9 × $10^{10}$ cells/ml, with the total event count $N = 5080$ (Methods); for panels d,e, the measurement was performed at cell density 8.1 × $10^{10}$ cells/ml, with the total event count $N = 3255$. Three independent experiments for each of panels b-e reproduced similar distributions. (f) Nematic alignment parameters $Q_{\text{cell}} = \langle 2\cos^2\varphi_{\text{cell}} - 1\rangle$ and $Q_{\text{fla}} = \langle 2\cos^2\varphi_{\text{fla}} - 1\rangle$ plotted against motile cell proportion $r_m$. See panel b for definition of the angles $\varphi_{\text{cell}}$ and $\varphi_{\text{fla}}$. Total cell density was maintained at ~8 × $10^{10}$ cells/ml. Error bars are standard deviations of 3 independent experiments and the total event count in each experiment $N = 15814 \pm 2149$ (mean±s.d.). All measurements in this figure were performed at the mid-plane of the microfluidic chamber.

## Supplementary Tables

| Name | Symbol | Notes |
|---|---|---|
| cell body center | | Point |
| cell body long axis | $b\mathbf{n}_{\mathrm{cell}}$ | $\mathbf{n}_{\mathrm{cell}} = -\mathbf{n}_{\mathrm{cell}}$ is a director |
| cell body short axis | $b'\mathbf{n}_{\mathrm{cell}}^{\perp}$ | $\mathbf{n}_{\mathrm{cell}}^{\perp}$ is a director and orthogonal to $\mathbf{n}_{\mathrm{cell}}$ |
| flagellum center | | Point |
| flagellum long axis | $f_1\mathbf{n}_{\mathrm{fla}}$ | $\mathbf{n}_{\mathrm{fla}} = -\mathbf{n}_{\mathrm{fla}}$ is a director |
| flagellum tail end | | Point |
| flagellum vertex | | Point |
| vector from flagellum center to cell body center | $a\mathbf{p}$ | $\mathbf{p}$ is a unit vector |
| vector from flagellum vertex to cell body center | $a_1\mathbf{p}_1$ | $\mathbf{p}_1$ is a unit vector |
| vector from flagellum tail to flagellum vertex | $a_2\mathbf{p}_2$ | $\mathbf{p}_2$ is a unit vector |
| solvent-flow velocity | $\mathbf{u}_{\mathrm{flow}}$ | 2D vector array |
| local solvent-flow velocity (interpolated) | $\mathbf{u}_{\mathrm{flow}}$ | Vector with magnitude $u_{\mathrm{flow}}$ |
| apparent single-cell velocity | $\mathbf{u}_{\mathrm{cell}}$ | Vector with magnitude $u_{\mathrm{cell}}$ |
| advection-corrected single-cell velocity | $\mathbf{v}_{\mathrm{cell}}$ | Vector with magnitude $v_{\mathrm{cell}}$ |

Table S1. Microscopic quantities related to the dynamics of cellular motion that can be measured experimentally.

| Angle | Symbol | Polar or nematic alignment parameter |
|---|---|---|
| angle between $\mathbf{u}_{\mathrm{cell}}$ and $\mathbf{p}$ | $\theta_{\mathrm{u}}$ | $\langle\cos\theta_{\mathrm{u}}\rangle$ |
| angle between $\mathbf{v}_{\mathrm{cell}}$ and $\mathbf{p}$ | $\theta_{\mathrm{v}}$ | $\langle\cos\theta_{\mathrm{v}}\rangle$<br>($\mathbf{v}_{\mathrm{cell}}$-$\mathbf{p}$ misalignment parameter: $D_{\mathrm{v}} = 1 - \langle\cos\theta_{\mathrm{v}}\rangle$) |
| angle between $\mathbf{u}_{\mathrm{flow}}$ and $\mathbf{p}$ | $\theta$ | $\langle\cos\theta\rangle$ |
| direction of $\mathbf{p}$ in the laboratory reference frame | $\theta'$ | $\langle\cos\theta'\rangle$ |
| angle between $\mathbf{u}_{\mathrm{cell}}$ and $\mathbf{u}_{\mathrm{flow}}$ | $\varphi_{\mathrm{u}}$ | $\langle\cos\varphi_{\mathrm{u}}\rangle$ |
| angle between $\mathbf{v}_{\mathrm{cell}}$ and $\mathbf{u}_{\mathrm{flow}}$ | $\varphi_{\mathrm{v}}$ | $\langle\cos\varphi_{\mathrm{v}}\rangle$<br>(tendency of upstream navigation: $-\langle\cos\varphi_{\mathrm{v}}\rangle$) |
| angle between $\mathbf{n}_{\mathrm{cell}}$ and $\mathbf{u}_{\mathrm{flow}}$ | $\varphi_{\mathrm{cell}}$ | $Q_{\mathrm{cell}}$ |
| angle between $\mathbf{n}_{\mathrm{fla}}$ and $\mathbf{u}_{\mathrm{flow}}$ | $\varphi_{\mathrm{fla}}$ | $Q_{\mathrm{fla}}$ |

Table S2. Angles related to the orientational dynamics of cellular motion.